\documentclass{vldb1}

\makeatletter
\def\@copyrightspace{\relax}
\makeatother

\usepackage{balance}
\usepackage{wrapfig}
\usepackage{graphicx}
\usepackage{balance}
\usepackage{setspace}
\usepackage{enumitem}
\usepackage{multirow}
\usepackage{hhline}

\usepackage{theorem}
\usepackage{wrapfig}
\usepackage{amssymb}
\usepackage{xspace}

\usepackage{color}

\usepackage[boxed,ruled,vlined,linesnumbered]{algorithm2e}
\usepackage{url}
\usepackage{framed}
\usepackage{hyperref}
\usepackage{caption}
\usepackage{subcaption}
\usepackage{times}
\usepackage{graphicx}
\newcommand{\B}{\vspace*{-\smallskipamount}}
\newcommand{\BB}{\vspace*{-\medskipamount}}
\newcommand{\BBB}{\vspace*{-\bigskipamount}}

\SetAlgoSkip{}
\usepackage{balance}

\usepackage{xcolor,colortbl}
\usepackage{listings}

\usepackage{tikz}
\newcommand\encircle[1]{%
\tikz[baseline=(X.base)]
   \node (X) [draw, shape=circle, inner sep=-1.5pt, fill=black, text=white] {\strut #1};}

\definecolor{codegreen}{rgb}{0,0.6,0}
\definecolor{codegray}{rgb}{0.5,0.5,0.5}
\definecolor{codepurple}{rgb}{0.58,0,0.82}
\definecolor{backcolour}{rgb}{0.95,0.95,0.92}

\lstdefinestyle{mystyle}{
  commentstyle=\color{codegreen},
  keywordstyle=\color{magenta},
  numberstyle=\tiny\color{codegray},
  stringstyle=\color{codepurple},
  basicstyle=\scriptsize,
  breakatwhitespace=false,
  breaklines=true,
  captionpos=b,
  keepspaces=true,
  numbersep=5pt,
  showspaces=false,
  showstringspaces=false,
  showtabs=false,
  tabsize=2,
  belowskip=-\baselineskip,
  aboveskip=-0.05cm
}

\vldbTitle{{\textsc{{\Huge Quest}}}: Practical and Oblivious Mitigation Strategies for COVID-19 using WiFi Datasets}
\vldbAuthors{Peeyush Gupta, Sharad Mehrotra, Nisha Panwar, Shantanu Sharma, Nalini Venkatasubramanian, and Guoxi Wang}
\vldbDOI{h}

\begin{document}


\title{{\textsc{{\huge Quest}}}: Practical and Oblivious Mitigation Strategies for COVID-19 using WiFi Datasets\thanks{{\scriptsize We are thankful to Dhrubajyoti Ghosh for helping us in the system installation. \newline This material is based on research sponsored by DARPA under agreement number FA8750-16-2-0021. The U.S. Government is authorized to reproduce and distribute reprints for Governmental purposes notwithstanding any copyright notation thereon. The views and conclusions contained herein are those of the authors and should not be interpreted as necessarily representing the official policies or endorsements, either expressed or implied, of DARPA or the U.S. Government. This work is partially supported by NSF grants 1527536 and 1545071.}}}



%
%
%
%

\numberofauthors{1} 

\author{Peeyush Gupta$^1$, Sharad Mehrotra$^1$, Nisha Panwar$^2$, Shantanu Sharma$^1$, \\ Nalini Venkatasubramanian$^1$, and Guoxi Wang$^1$
\\$^1$University of California, Irvine, USA. $^2$Augusta University, USA. \\ Email: \texttt{sharad@ics.uci.edu, shantanu.sharma@uci.edu}
}

\maketitle
\BBB\BBB
\begin{abstract}
Contact tracing has emerged as one of the main mitigation strategies to prevent the spread of pandemics such as COVID-19. Recently, several efforts have been initiated  to track individuals, their movements, and interactions using technologies, \textit{e}.\textit{g.}, Bluetooth beacons, cellular data records, and smartphone applications. Such solutions are often intrusive,  potentially violating individual privacy rights and are often subject to regulations (\textit{e}.\textit{g.}, GDPR and CCPR) that mandate the need for opt-in policies to gather and use  personal information. In this paper, we introduce \textsc{Quest}, a system that empowers organizations to observe individuals and spaces to implement policies for social distancing and contact tracing using WiFi connectivity data in a passive and {\bf \em privacy-preserving manner}. The goal is to ensure the safety of  employees and occupants at an organization, while protecting the privacy of all parties. \textsc{Quest} incorporates computationally- and information-theoretically-secure protocols that prevent adversaries from gaining knowledge of an individual's location history (based on WiFi data); it includes support for accurately identifying users who were in the vicinity of a confirmed patient, and then informing them via opt-in mechanisms. \textsc{Quest} supports a range of privacy-enabled applications to ensure adherence to  social distancing, monitor the flow of people through spaces, identify potentially impacted regions, and raise exposure alerts. We describe the architecture, design choices, and implementation of the proposed security/privacy techniques in \textsc{Quest}. We, also, validate the practicality of \textsc{Quest} and evaluate it thoroughly via an actual campus-scale deployment at UC Irvine over a very large dataset of over 50M tuples.
\end{abstract}

\subsection*{Keywords} COVID-19, contact tracing, location tracing, social distancing, crowd-flow, WiFi.

\section{Introduction}
\label{sec:introduction}
The ongoing COVID-19 pandemic with rapid and widespread global impact, has caused havoc over the past few months --- at the time of writing of this paper, over 3 million individuals have been infected.   The epidemic  has caused over  200,000 global casualties, and the world economy to come to a screeching halt. Several (non-pharmacologic) steps are being taken by governments and organizations to restrict the spread of the virus, including social distancing measures, quarantining of those with confirmed cases, lock-down of non-essential businesses, and contact-tracing methods to identify and warn potentially exposed individuals. These tracking and tracing measures utilize a range of technological solutions. Countries, \textit{e}.\textit{g}., Israel, Singapore, China, Taiwan, and Australia, utilize cellular data records or data from Bluetooth-enabled apps to perform contact tracing.  Other countries, \textit{e}.\textit{g}.,  India,  have begun manual contact tracing by interviewing patients.

Recently, commercial and academic solutions (\textit{e}.\textit{g}., Apple-Google collaboration~\cite{google-apple}, European PEPP-PT~\cite{pepppt}, Israel's The Shield~\cite{IsraelsTheShield}, Singapore's TraceTogether~\cite{TraceTogether}, South Korea's 100m~\cite{SouthKorea100m}, and~\cite{DBLP:journals/corr/abs-2004-05251,DBLP:journals/corr/abs-2004-04145,DBLP:journals/corr/abs-2003-13073,canetti2020anonymous,epfl})  aim to provide secure contact tracing using Bluetooth-based proximity-detection. Using this approach, users can check if they have been exposed to a potential carrier of the virus by performing a private set intersection of their data with the secured public registry of infected people. While this approach is a step towards protecting the privacy of individuals, there are several limitations:
First, the collection and sharing of such personal information can compromise the privacy of individuals --- there are growing fears that this could also lead to misuse of data (now or in the future), \textit{e}.\textit{g}., mass surveillance of communities and targeting of specific populations~\cite{cho2020contact,epfl}.
Second, such methods require users to opt-in to broadcast, share, and collect the data using Bluetooth --- past work has highlighted limited adoption of such technologies, especially, in parts of the world where privacy is considered to be a paramount  concern~\cite{DBLP:journals/corr/abs-2004-05251,DBLP:conf/percom/BiXHHPM17}.
Third, contact tracing using Bluetooth or GPS-based proximity sensing has been shown to have false positives/negatives, leading to limited accuracy~\cite{canetti2020anonymous,DBLP:journals/corr/abs-2004-05251}.
Finally, past experiences have indicated that creating pathways for large organizations to capture personal data can lead to data theft, \textit{e}.\textit{g}.,  Facebook's Cambridge Analytica situation and~\cite{DBLP:journals/corr/abs-1803-09007}.

Contact tracing approaches are reactive in nature and aim to detect exposure after it occurs. We argue that proactive and preventive approaches are critical to contain and mitigate the spread. For instance, the ability to monitor public spaces (\textit{e}.\textit{g}., classrooms, restaurants, malls), which are expected to have significant density and population flow, can be used by organizations (\textit{e}.\textit{g}., campuses) to observe the extent to which employees (and employers) are adhering to social distancing directives. In fact, based on recent media articles~\cite{cnn,vox} and conversations with our university leadership,\footnote{{\scriptsize Including epidemiologists in the public health school.}}
the importance of such applications will increase further as organizations consider ways forward to reopen and resume operations. Today, organizations are working to help strike the right balance between onsite/online operations that afford both business continuity and public safety.

This paper describes our proposed solution, entitled \textsc{Quest} that exploits existing WiFi infrastructure (prevalent in almost every modern organization) to support a sleuth of applications that empower organizations to evaluate and tune directives for safe operation, while protecting the privacy of the individuals in their premises. Particularly, \textsc{Quest} leverages WiFi connectivity data (the data generated when a device connects to wireless access-points, see \S\ref{sec:overview of quest} for details) to support applications for social distancing adherence, crowd-flow, contact tracing, and exposure notifications within  premises (both inside/outside buildings). The WiFi data collected is appropriately secured to prevent leakage of personally identifiable information (\textit{e}.\textit{g}., MAC address of the mobile device) and outsourced to a public (cloud) server. On the outsourced data, \textsc{Quest} allows application execution to occur in a privacy-preserving manner.

\textsc{Quest} supports two different cryptographic alternatives for secure data processing; the choice of the approach is based on underlying security requirements of the organization. The first is a computationally secure encryption-based mechanism, entitled \textsc{cQuest} that encrypts data using a variant of  searchable encryption methods. The second approach called \textsc{iQuest}, based on a string-matching technique~\cite{DBLP:conf/ccs/DolevGL15} over secret-shares generated using Shamir's secret-sharing algorithm~\cite{DBLP:journals/cacm/Shamir79}. Both methods support the above-mentioned applications.
\textsc{iQuest} offers a higher level of security, when using untrusted servers, since it is information-theoretically secure, and moreover, does not reveal access-patterns (\textit{i}.\textit{e}., the identity of tuples satisfying the query). We have deployed \textsc{Quest} at UC Irvine~\cite{uciapp}, as well as, tested the system on large WiFi datasets. These datasets were collected as a part of the TIPPERS smartspace testbed at UCI~\cite{DBLP:conf/percom/MehrotraKVR16} and will also be used for scalability studies.

\textsc{Quest} offers several  distinct advantages compared to other ongoing contact tracing efforts that have focused on using GPS, cellular infrastructure, and proximity sensors (\textit{e}.\textit{g}., Bluetooth)~\cite{stanford,TraceTogether,google-apple,canetti2020anonymous,DBLP:journals/corr/abs-2003-13073,DBLP:journals/corr/abs-2004-05251}. These include:

\begin{itemize}[leftmargin=0.01in]
\item {\em A Decentralized organizational solution.} \textsc{Quest} is designed as a tool to be used independently and autonomously by organizations (\textit{e}.\textit{g}., universities, individual shops/shopping complexes, and airports) to monitor adherence of their policies for social distancing, crowd-flow, and, to warn people about possible exposure on their premise. The organizational aspect of \textsc{Quest} brings several advantages. First, the solution is amenable for organizational-level control to ensure that warning and alerts are not misused to spread false information, unlike some of the recent tools which are being targeted by malicious adversaries to spread propaganda and misinformation~\cite{fakenews1,fakenews2}. Second, unlike solutions such as the one being designed by mobile OS platform vendors (viz. Apple and Google), in \textsc{Quest}, both data collection and usages remain decentralized to the level of an organization and, thus, end-users do not need to trust any single organization/authority with their data.

\item \emph{A robust solution that works both inside buildings and outdoors.} Since \textsc{Quest} is based on WiFi technology, it has a distinct advantage of being able to monitor both inside buildings (organizational premises) and in outside spaces, due to the ubiquitous nature of WiFi coverage in both indoors and outdoors of campuses. The use of WiFi data brings in several additional advantages: First, \textsc{Quest} does not require any additional hardware expenses or deployment of any new technology that might be prohibitive in terms of cost and limited in terms of deployment. Second, since WiFi connectivity events are generated automatically by current WiFi protocols, \textsc{Quest} is entirely passive, \textit{i}.\textit{e}., it does not require users to deploy any new applications or make changes to their mobile devices. Third, the technology is platform independent, since data collection is implemented entirely on the infrastructure side.

\item \emph{Privacy-by-design.} \textsc{Quest} supports the above-mentioned applications in a privacy-preserving manner by exploiting computationally secure and information-theoretically secure techniques. Thus, \textsc{Quest} does not provide additional information about people, their locations, or their health status to any organization that they do not already have. Also, an adversary cannot learn  past behavior or predict the future behavior of any user. Since the ciphertext representations of a person across organizations are different, even from jointly observing data of multiple organizations to know any specific person has been to the premises of one or more organizations.\footnote{{\scriptsize Organizations today, if they so desire, can capture and trace individuals based on their WiFi connectivity data. \textsc{Quest}, obviously, cannot prevent such a use of WiFi data. The key-point is that while \textsc{Quest} stores secured WiFi data at the cloud, the data-at-rest or query execution will not reveal any additional information to the organizations.}}

\end{itemize}

\noindent\textbf{Outline.} \S\ref{sec:Preliminary} provides the model and security properties. \S\ref{sec:overview of quest} provides an overview of \textsc{Quest} and its applications. \S\ref{sec:Computationally-Secure Solution} provides \textsc{cQuest} protocol. \S\ref{sec:Information-Theoretically Secure Solution} provides \textsc{iQuest} protocol. We evaluate \textsc{Quest} in \S\ref{sec:Experimental Evaluation} and compare it with other state-of-the-art approaches, \textit{e}.\textit{g}., Intel Software Guard Extensions (SGX)~\cite{sgx} based Opaque~\cite{opaque} and multi-party computation (MPC)-based Jana~\cite{jana};
we discuss tradeoffs between security and performance.

\section{Related Work and Comparison}
\label{sec:Comparison With Existing Work}

In this section, we discuss the new approaches designed for COVID-19 contact tracing, several prior research approaches have explored proximity-based solutions to monitor the spread of infections, and compare against \textsc{Quest}.

\medskip
\noindent
\textbf{Comparison with COVID-19 proximity finding approaches.}
Several recent approaches for preventing the spread of coronavirus are based on Bluetooth data-based secure proximity detection. For example, Canetti et al.~\cite{canetti2020anonymous} present a person proximity detection method based on Bluetooth-enabled devices. However, this method requires to store parts of the data at the user device. ~\cite{canetti2020anonymous}, also, argued that GPS-based proximity detection inside a building can give false results. Stanford University~\cite{stanford} is also developing applications based on Bluetooth data. Singapore's TraceTogether application~\cite{TraceTogether}, also, works based on Bluetooth-based tracking. However,~\cite{cho2020contact} showed that TraceTogether jeopardizes the user privacy. DP-3T (decentralized privacy-preserving proximity tracing)~\cite{epfl} proposed a proximity tracing system based on Bluetooth data. Google and Apple~\cite{google-apple} are developing Bluetooth beacon-based contact tracing, while preserving the user privacy and location privacy. Similar work is also proposed in~\cite{DBLP:journals/corr/abs-2003-13073,DBLP:journals/corr/abs-2004-05251} for Bluetooth data-based secure proximity detection, based on the private set intersection.  Enigma MPC, Inc.~\cite{enigma} develops SafeTrace that requires users to send their encrypted Google Map timeline to a server equipped with Intel SGX~\cite{sgx} that executes contact tracing and finds whether the person got in contact with an impacted person or not.
A survey of recent contact tracing application for COVID-19 may be found in~\cite{DBLP:journals/corr/abs-2004-06818}. However, all such methods require either to install an application~\cite{epfl,TraceTogether} at the device, to store some data~\cite{canetti2020anonymous,DBLP:journals/corr/abs-2003-13073} at the device, to execute  computation~\cite{canetti2020anonymous,enigma,DBLP:journals/corr/abs-2003-13073} at the device, to explicitly opt-in to enable Bluetooth-based beacon exchange~\cite{google-apple,DBLP:journals/corr/abs-2003-13073,DBLP:journals/corr/abs-2004-05251}, or jeopardize the user privacy~\cite{cho2020contact}.

In contrast, \textsc{Quest} does not require any effort by users, since we rely on WiFi data that is generated when a device connects with a WiFi network. \textsc{Quest} discovers the most accurate proximity of a person inside a building, unlike GPS-based approaches. Also, while using the servers, \textsc{iQuest} provides complete security, due to using secret-sharing based technique. \textsc{Quest} not only provides contact tracing, but also provides other applications (Table~\ref{tab:queries}).

\medskip
\noindent
\textbf{Comparison with other proximity finding approaches.}
Epic~\cite{DBLP:conf/icc/AltuwaiyanHL18} and Enact~\cite{DBLP:conf/mobisys/PrasadK17} are based on WiFi signal strength, where a dynamic user scans the surrounding's wireless signals, access-points, and records in their phones. The infected user sends this information to a server that notifies other users and requests them to find their chances of contact. However, Epic~\cite{DBLP:conf/icc/AltuwaiyanHL18} and Enact~\cite{DBLP:conf/mobisys/PrasadK17} consider trust in reporting by the infected users and requires storing some information at the smartphone, like Bluetooth-based solutions~\cite{epfl,TraceTogether,canetti2020anonymous,enigma,DBLP:journals/corr/abs-2003-13073}. Another problem with such signal strength-based methods is in developing models to compare WiFi signals and have issues related to spatial, temporal, and infrastructural sensing~\cite{kjaergaard2012challenges}. NearMe~\cite{DBLP:conf/huc/KrummH04}, ProbeTag~\cite{DBLP:conf/wimob/MaierSD15},~\cite{DBLP:journals/imwut/SapiezynskiSWLL17},~\cite{meunier2004peer}, and~\cite{krumm2009proximity} proposed  similar approaches for proximity detection. The seminal work~\cite{DBLP:conf/eurocrypt/BrandsC93} proposed distance-bounding protocol to estimate an upper-bound on the physical proximity of the device through the round-trip time measurements, by exchanging unique challenge-response pairs between a sender and a receiver.~\cite{DBLP:conf/icalp/GellesOW12} provided a solution for proximity testing among the users while hiding their locations by encryption and considered user-to-user- and server-based proximity testing. Note that all such methods require \emph{active participation} from the users.

In contrast, \textsc{Quest} does not require active participation from the user, since it relies on WiFi connectivity data, which is, obviously, generated when a device gets connected with a WiFi network.

\medskip
\noindent
\textbf{Background on cryptographic techniques.}
We may broadly classify existing cryptographic techniques into two categories:
(\textit{i}) \emph{Computationally secure solutions} that includes encryption-based techniques such as symmetric-searchable encryption (SSE)~\cite{DBLP:conf/sp/SongWP00,DBLP:journals/jcs/CurtmolaGKO11,DBLP:journals/pvldb/LiLWB14,DBLP:conf/icde/LiL17}, deterministic encryption~\cite{DBLP:conf/crypto/BellareBO07,DBLP:conf/crypto/BoldyrevaFO08}, and order-preserving encryption (OPE)~\cite{DBLP:conf/sigmod/AgrawalKSX04},
(\textit{ii}) \emph{information-theoretically secure solutions} that include secret-sharing-based techniques~\cite{DBLP:journals/cacm/Shamir79,DBLP:conf/ccs/DolevGL15} and multi-party computation (MPC) techniques~\cite{jana}.
Computationally secure solutions, such as SSE --- PB-tree~\cite{DBLP:journals/pvldb/LiLWB14} and IB-tree~\cite{DBLP:conf/icde/LiL17}, are efficient in terms of computational time. However, they
(\textit{i}) reveal access-patterns (\textit{i}.\textit{e}., the identity of the tuple satisfying the query),
(\textit{ii}) do not scale to a large-dataset due to dependence of a specific index structure,
(\textit{iii}) are not efficient for \emph{frequent data insertion}, since it requires to rebuild the entire index at the trusted side, and
(\textit{iv}) cannot protect data from a computationally-efficient adversary or the government legislation/subpoena that may force to give them the data in cleartext. In contrast, information-theoretically secure solutions hide access-patterns, as well as, secure against a computationally-efficient adversary or the government legislation/subpoena, (if the shares of the data are placed at the public servers under the different jurisdiction). Instead of using any cryptographic solution, one may also use \emph{secure hardware-based solutions} that include Intel Software Guard eXtension (SGX)~\cite{sgx} based systems, \textit{e}.\textit{g}., Opaque~\cite{opaque}, Bunker and Fort~\cite{DBLP:conf/eurosec/AmjadKM19},  HardIDX~\cite{DBLP:journals/jcs/FuhryBBHKS18}, and EncDBDB~\cite{DBLP:journals/corr/abs-2002-05097}. However, such solutions suffer from similar issues as computationally secure solutions and suffer from additional side-channel attacks, such as cache-line~\cite{DBLP:conf/eurosec/GotzfriedESM17} and branching attacks~\cite{DBLP:conf/ccs/WangCPZWBTG17} that reveals access-patterns.

\section{Preliminary}
\label{sec:Preliminary}
This section explains the entities involved in deploying \textsc{Quest}, the adversarial model, and the desired security properties.

\newpage
\subsection{Entities}
We have the following two major entities in \textsc{Quest}. 
\begin{enumerate}[leftmargin=0.01in]
\item An \emph{organization} $o_i$, who owns and deploys WiFi infrastructure (\textit{e}.\textit{g}., WiFi access-points), and hosts \textsc{Quest} that receives WiFi (connectivity) data (from the infrastructure) of the form: $\langle d_i,l_i,t_i\rangle$, where $d_i$ is the $i^{\mathit{th}}$ device-id and $t_i$ is the time when $d_i$ connects with a WiFi access-point $l_i$. Prior to outputting the data, \textsc{Quest} appropriately implements a cryptographic technique to prevent misuse of the data from an adversary.

\item The \emph{untrusted public (cloud) servers} that host the secured data, outsourced by \textsc{Quest}, on which they execute applications. We assume that the servers support any database system, \textit{e}.\textit{g}., MySQL.

\end{enumerate}

Also, we assume two additional entities: a \emph{querier} and a \emph{publisher} ($\mathcal{P}$). A \emph{querier} (which may be the organization or any (authenticated) person) is allowed to execute \textsc{Quest} applications on the secured data (via \textsc{Quest}).  Further, only for \emph{contact tracing application}, we assume a \emph{trusted publisher} $\mathcal{P}$ (\textit{i}.\textit{e}., hospitals), who publishes a secured list $\mathcal{L}$ of device-ids of confirmed infected person.
We assume that \textsc{Quest} executes a secure authentication protocol with $\mathcal{P}$ to confirm the queried device-id as the device-id of an infected person, before executing contact tracing application. Thus, it prevents the querier to execute a query for any device-id.

\subsection{Adversarial Model}
\label{subsec:Security Properties}
As we have two entities in \textsc{Quest}. Below, we discuss their adversarial behavior and our assumptions.

\smallskip
\noindent\textbf{\emph{Organizations.}} We assume that organizations have their own security keys (\textit{e}.\textit{g}., public and private keys). A \emph{system that hosts \textsc{Quest}, its security keys, and programs must be secured and untampered by anyone} including the organizations, and this is an inevitable assumption, as well as, similar to the database-as-a-service (DaS) model~\cite{DBLP:conf/icde/HacigumusMI02}, (where it is assumed that an adversary cannot filtrate the information stored at the database owners). Otherwise, any cryptographic mechanism cannot be implemented on WiFi data. Also, it is important to mention that $o_i$ can capture their infrastructure data and can do any computation on the data as per their desire, without informing anyone. Controlling such organizations and preventing the misuse of the data is not in the scope of \textsc{Quest}. Recall that \textsc{Quest}'s goal is to prevent $o_i$ to track individuals or to run any applications other than those supported by \textsc{Quest} on the data collected by \textsc{Quest}.

\smallskip
\noindent\textbf{\emph{Cloud servers.}} We assume that the public servers are honest-but-curious (HBC) and/or malicious adversaries. Such adversarial models are considered widely in data outsourcing techniques~\cite{DBLP:conf/icde/HacigumusMI02,DBLP:journals/jcs/CurtmolaGKO11,DBLP:conf/stoc/CanettiFGN96,DBLP:conf/ctrsa/IshaiKLO16,DBLP:conf/ccs/DolevGL15,DBLP:conf/icdcs/WangCLRL10,DBLP:conf/ccs/YuWRL10}. An HBC adversary may wish to learn the user information by observing query execution, while a malicious adversary deviates from the algorithm. Since dealing with a malicious adversary, we use an information-theoretically secure solution that uses Shamir's secret-sharing~\cite{DBLP:journals/cacm/Shamir79}. Thus, we follow the restriction of Shamir's secret-sharing that the adversary cannot collude with all (or possibly the majority of) the servers.
Prior to sending the data to the server, we assume that \textsc{Quest} authenticates the servers. Also, we assume that the secret-shared data transmission to the servers is done using an anonymous routing protocol~\cite{DBLP:journals/cacm/GoldschlagRS99}, and it prevents an adversary to eavesdrop on a majority of communication channels between \textsc{Quest} and the servers, and thus, preventing the adversary to know the servers that store the secret-shares.

\begin{figure*}[!t]
    \centering
    \includegraphics[scale=0.7]{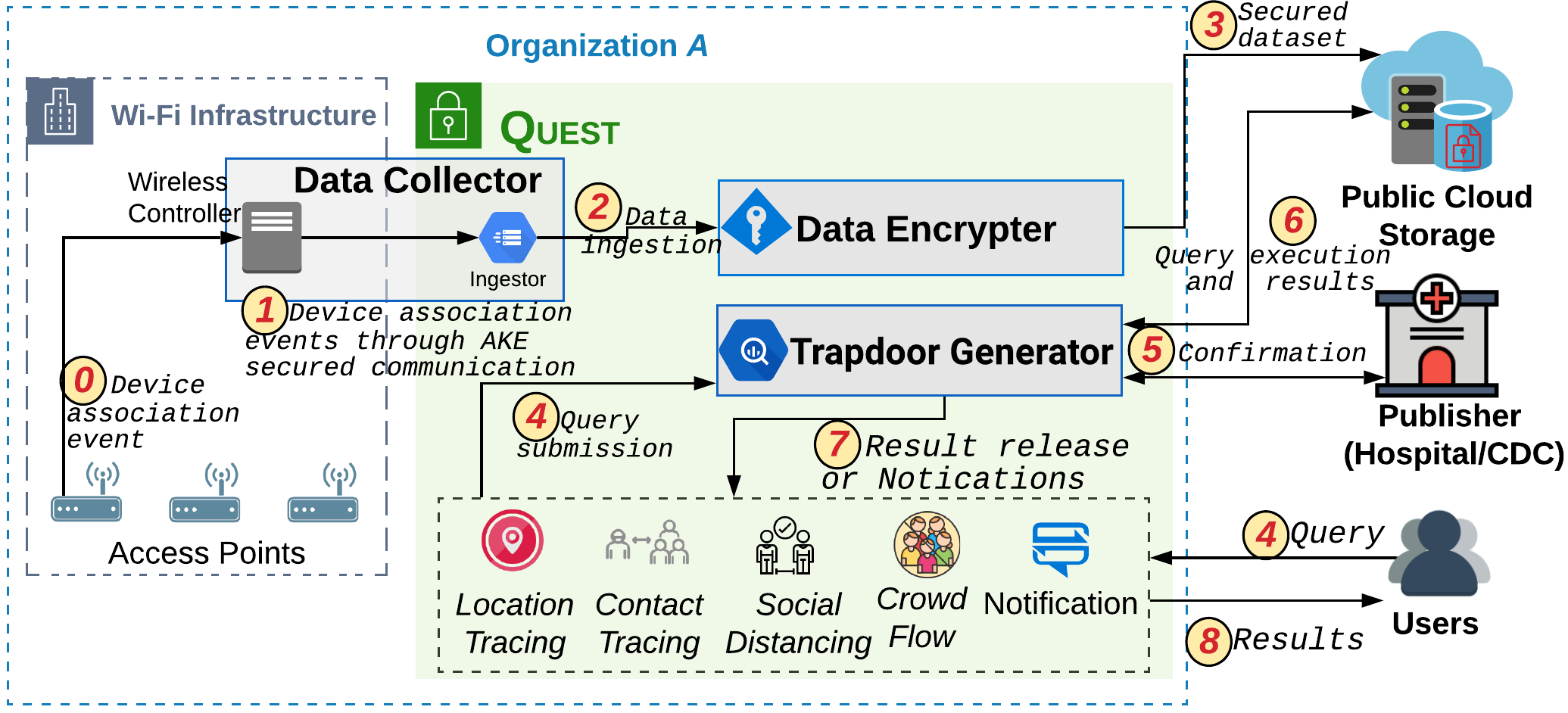}
    \caption{\textsc{Quest} system.}
    \label{fig:quest}
\BBB
\end{figure*}


\newpage
\subsection{Security Properties}
\label{subsec:Security Properties}
In the above-mentioned adversarial model, an adversary wishes to reveal user privacy by learning from data-at-rest and query execution. Thus, a secure algorithm must prevent an adversary to learn the data by just observing (\textit{ii}) cryptographically-secure data  and (\textit{ii}) query execution and deduce which tuples satisfy the query (\textit{i}.\textit{e}., access-patterns). Also, we need to ensure that a querier cannot execute a query for device-ids not published by $\mathcal{P}$. Thus, we need to maintain the following properties:

\smallskip
\noindent
\textbf{Ciphertext indistinguishability.} In the proposed scheme, the data contains user device-id. Thus, the  \emph{indistinguishability} of the user device-ids and locations is a vital requirement. Thus, the adversary, just by observing the secured dataset, cannot deduce that any two tuples belong to the same user/location or not. Note that satisfying indistinguishability property also prevents the adversary from learning any information from jointly observing two datasets belonging to two different organizations. 

\smallskip
\noindent
\textbf{Secure query execution.} It requires to maintain:
(\textit{i}) \emph{Query privacy} that prevents the adversary from distinguishing between two query predicates (for the same or different device-ids and locations) by observing the query predicates or by observing the two queries' execution, \textit{i}.\textit{e}., access-patterns.
(\emph{ii}) \emph{Execution privacy} that enforces the adversary to behave identically and to provide an identical answer to the same query. (Since an adversary cannot distinguish between two query predicates, it should follow the same protocol for each query execution to prove its non-adversarial behavior.) 

Satisfying these two properties achieve indistinguishability property during data-at-rest/query execution and do not reveal any information about the device-ids/locations. We can, formally, define it using the algorithm's real execution at the servers against the algorithm's ideal execution at a trusted party having the same data and the same query predicate. An algorithm reveals nothing if the real and ideal executions of the algorithm return the same answer.

\medskip
\noindent\textbf{Definition: Query privacy.} \emph{For any probabilistic polynomial time (PPT) adversary having a secured relation and any two input query predicates, say $p_1$ and $p_2$, the adversary cannot distinguish $p_1$ or $p_2$, either by observing the query predicates or by query output.}

\medskip
\noindent\textbf{Definition: Execution privacy.} \emph{For any given secured relation, any query predicate $\mathit{p}$ issued by any real user $U$, there exists a PPT user $U^{\prime}$ in the ideal execution, such that the outputs to $U$ and $U^{\prime}$ for the query predicate $\mathit{p}$ on the secured data are identical.}


\medskip
Note that satisfying the above two requirements (which are widely considered in many cryptographic approaches~\cite{DBLP:conf/stoc/CanettiFGN96,DBLP:conf/ctrsa/IshaiKLO16,DBLP:conf/ccs/DolevGL15}) 
will hide access-patterns, thus, the adversary cannot distinguish two different queries and the satisfying output tuples. However, such a secure algorithm (as given in \S\ref{sec:Information-Theoretically Secure Solution}) incurs the overhead. Thus, we relax the access-pattern-hiding property (similar to existing searchable encryption or secure-hardware-based algorithms) and, also, present efficient access-pattern revealing algorithm, \textsc{cQuest} (\S\ref{sec:Computationally-Secure Solution}).
In Appendix~\ref{app_sec:Security Property for Access-Pattern-Revealing Solutions}, we provide security property of \textsc{cQuest}.

\section{\textnormal{{\large\textbf{\textsc{Quest}}}} Architecture}
\label{sec:overview of quest}
\textsc{Quest}  contains the following three major components (see Figure~\ref{fig:quest}):

\smallskip
\noindent
\textbf{Data collector.} It collects WiFi connectivity (or association event) data of form $\langle d_i,l_j,t_k\rangle$, when a device $d_i$ connects to a WiFi access-point (AP) $l_j$ at time $t_k$.
Particularly, at the infrastructure side, the collector contains a wireless controller that receives WiFi data from several APs (\encircle{0}), via several methods, \textit{e}.\textit{g}., SNMP (Simple Network Management Protocol) traps~\cite{schlener2001flexible,DBLP:conf/huc/ZhouMZSPM16}, recent network management protocol NETCONF~\cite{enns2006netconf}, or Syslog~\cite{gerhards2009rfc}) and forwards WiFi data to \textsc{Quest} (\encircle{1}) over the network using the secure networking protocol (\textit{e}.\textit{g}., SSH~\cite{ylonen2006secure}). 
\textsc{Quest} receives and handles a large amount of streaming WiFi data at a very high rate (\encircle{1}). However, the encrypter may not be able to handle a sudden burst of data, due to the overhead of security techniques and may drop some data. Thus, \textsc{Quest} data collector includes an ingester (\textit{e}.\textit{g}., Apache Kafka, Storm, and Flume) that acts as a buffer between the wireless controller and the encrypter (\encircle{2}).

\smallskip
\noindent
\textbf{Data encrypter.} The encrypter collects data for a fixed interval duration, called \emph{epoch} (the reason of creating epochs will be clear soon in \S\ref{sec:Computationally-Secure Solution}) and then implements a cryptographic technique (based on the desired security level, using \textsc{cQuest} Algorithm~\ref{alg:Data Encryption Algorithm} or \textsc{iQuest} ~\ref{alg:Secret-share creation algorithm}) and outputs the secured data that is outsourced to the servers (\encircle{3}).

\smallskip\noindent
\textbf{Trapdoor generator.} A query/application is submitted to the trapdoor generator (\encircle{4}) that generates the secure trapdoor (using Algorithm~\ref{alg:Query Execution Algorithm} or~\ref{alg:iquest_Query Execution Algorithm_sss}) for query execution on secured data. For (user) contact tracing, it confirms the submitted device-id as the real device-id of an infected person or not, from the publisher (\encircle{5}). The trapdoors are sent to the servers that execute queries and send back encrypted results (\encircle{6}). The results are decrypted before producing the final answer (\encircle{7}). Further, the organization may alert the users appropriately (via emails or phones), if devices have allowed the organization to inform about it, during their registration (\encircle{8}).

\begin{table*}[!t]
\B
    \centering
    \scriptsize
    \begin{tabular}{|l|p{15.3cm}|}\hline
 \textbf{Applications} & \textbf{SQL syntax} \\\hline
Location tracing & \begin{lstlisting}[language=SQL,mathescape=true]
SELECT DISTINCT locationId
FROM WiFiData
INNER JOIN InfectedUsers ON WiFiData.macId = InfectedUsers.macId
WHERE timestamp > $t_1$ AND timestamp < $t_2$\end{lstlisting}\\
\hline

Contact tracing  &
\begin{lstlisting}[language=SQL, mathescape=true]
SELECT DISTINCT WifiData.macId
FROM WiFiData LEFT OUTER JOIN InfectedUsers ON WiFiData.macID = InfectedUsers.macId
    (SELECT locationId, timestamp
     FROM WiFiData
     INNER JOIN InfectedUsers ON WiFiData.macId = InfectedUsers.macId
     WHERE timestamp > $t_1$ AND timestamp < $t_2$) AS InfectedLocations
WHERE WiFiData.locationId = InfectedLocations.locationId
AND EXTRACT(WiFiData.timestamp, $\Delta$) = EXTRACT(InfectedLocations.timestamp, $\Delta$)
AND InfectedUsers.macId IS NULL
\end{lstlisting}
\\ \hline
Social distancing     & \begin{lstlisting}[language=SQL,mathescape=true]
SELECT DISTINCT COUNT(MacId)
FROM WiFiData, 	
     (SELECT WiFiData.locationId, timestamp, COUNT(DISTINCT MacId)/capacity AS socialDistancing
      FROM WiFiData INNER JOIN Location ON WiFiData.locationId = Location.locationId
      WHERE timestamp > $t_1$ AND timestamp < $t_2$
      GROUP BY WiFiData.locationId, timestamp/300
      HAVING socialDistancing > maxAllowed}) AS Violations
WHERE WiFiData.locationId = Violations.LocationId
AND EXTRACT(WifData.timestamp, delta) = EXTRACT(Violations.timestamp, delta)
\end{lstlisting}
        \\\hline
 Crowd-flow & \begin{lstlisting}[language=SQL,mathescape=true]
SELECT locationId, COUNT(DISTINCT macId) AS usersVisited
FROM WiFiData
WHERE timestamp > $t_1$ AND timestamp < $t_2$
GROUP BY locationId
ORDER BY usersVisited DESC
LIMIT K
\end{lstlisting}

\\\hline
    \end{tabular}
    \caption{A sample of supported applications by \textsc{Quest} in SQL.}
    \label{tab:queries}
\end{table*}

\begin{figure}[!t]
\BBB
  \centering
  \includegraphics[scale=0.13]{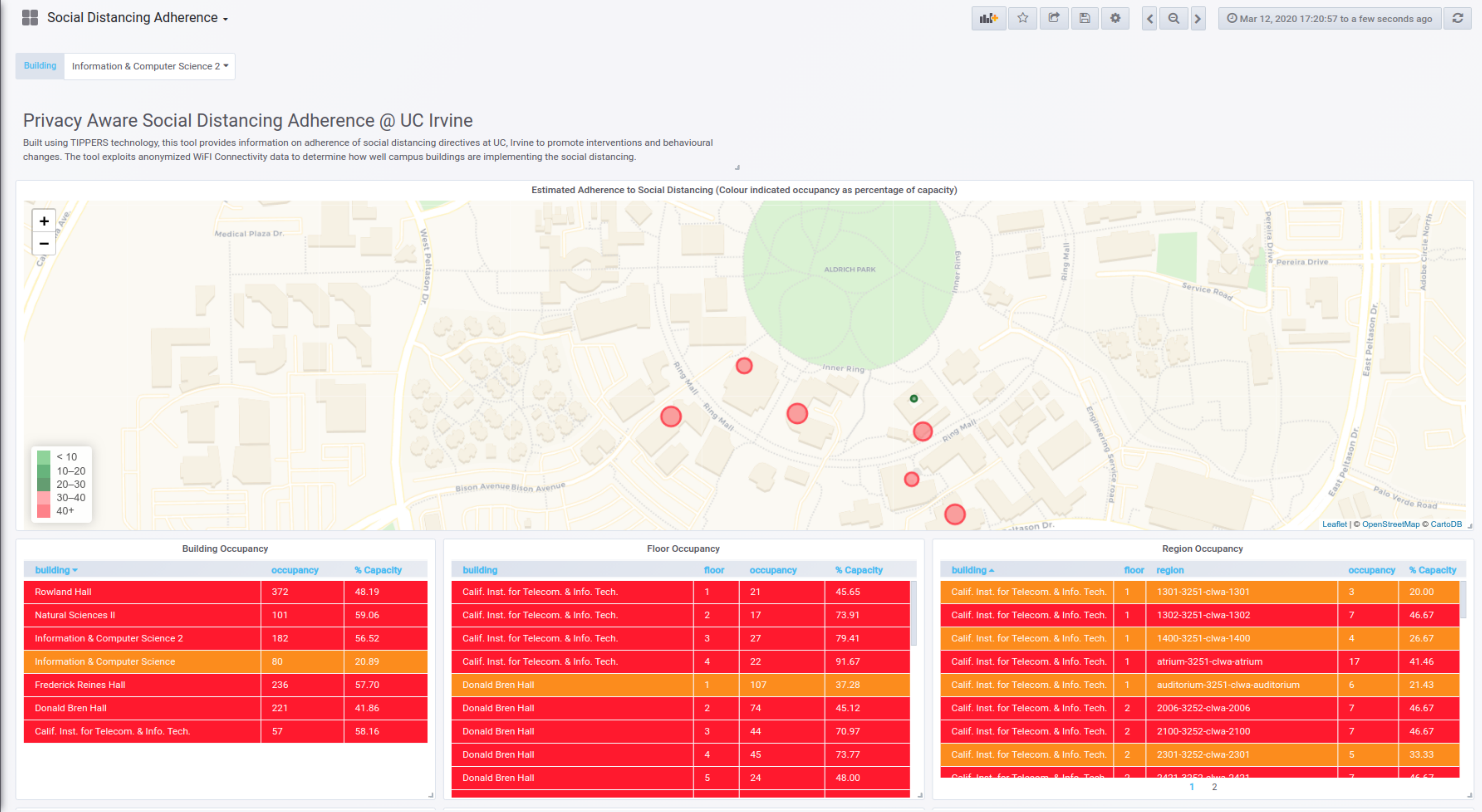}
  \caption{Social distancing application interface before lockdown.}
  \label{fig:Social distancing application interface before lockdown.}
\end{figure}

\begin{figure}[!t]
  \centering
  \includegraphics[scale=0.13]{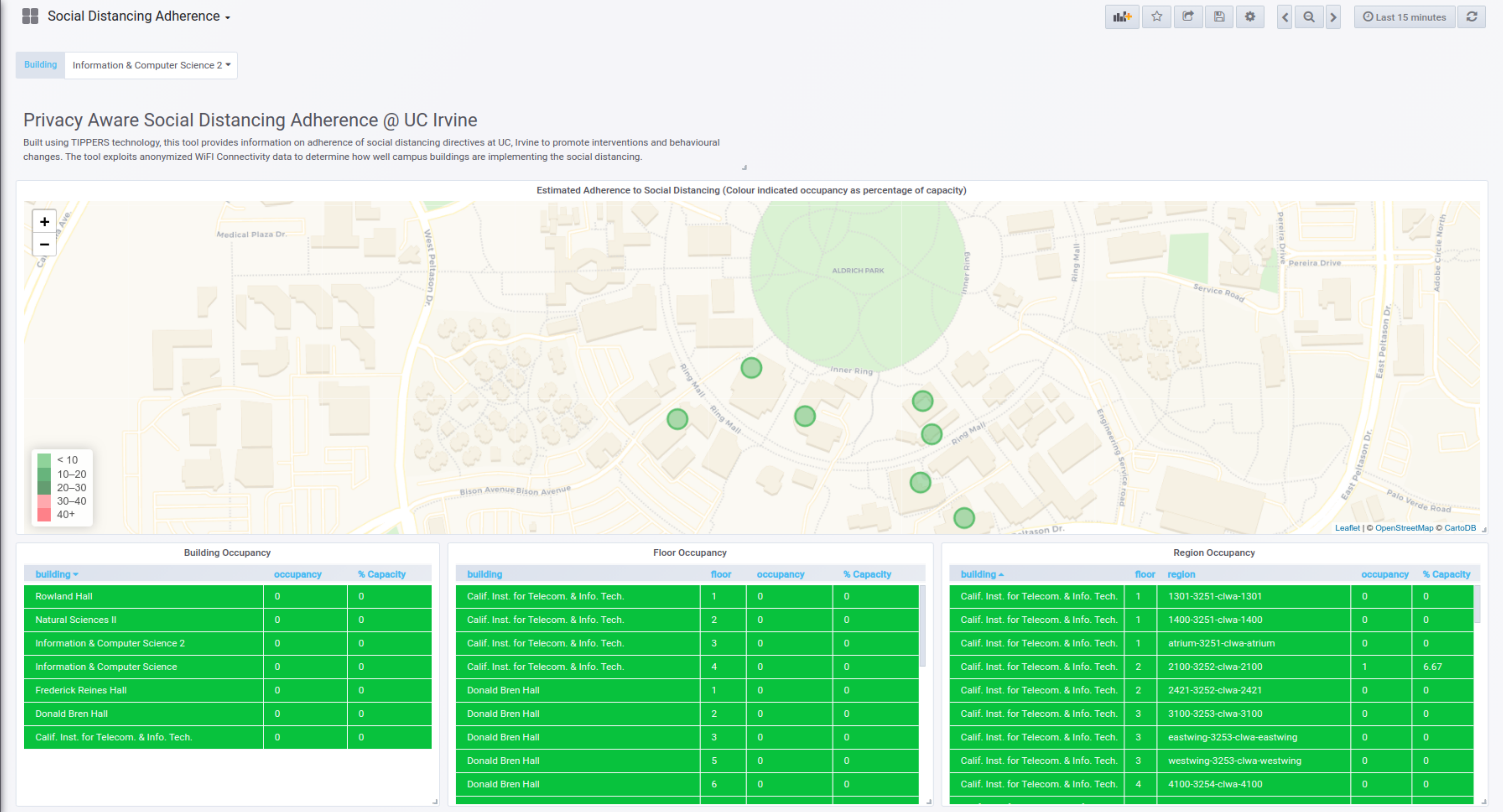}
  \caption{Social distancing application interface after lockdown.}
  \label{fig:Social distancing application interface after lockdown.}
\BBB
\end{figure}

\begin{figure}[!t]
\BBB
  \centering
  \includegraphics[scale=0.13]{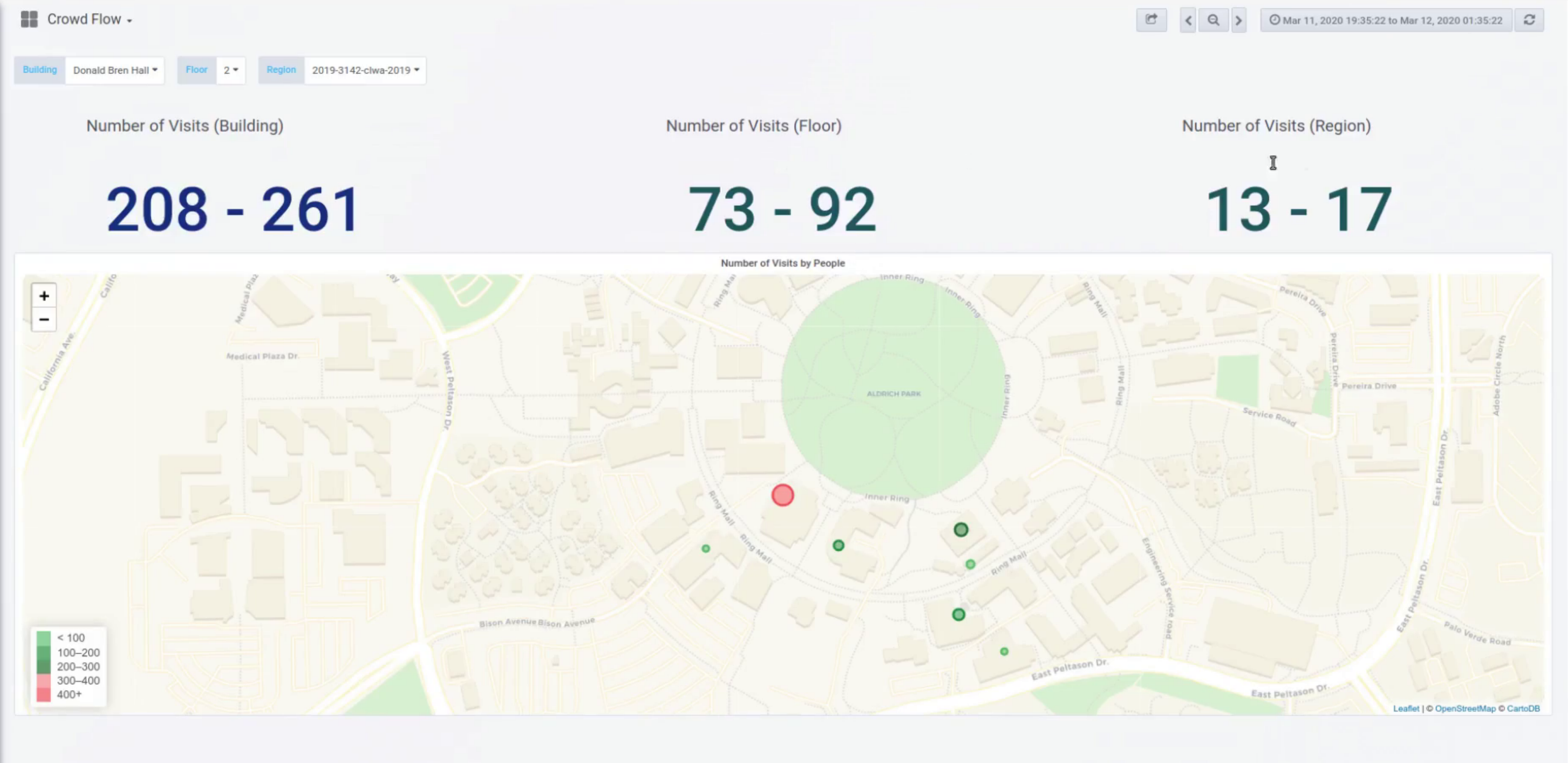}
  \caption{Crowd-flow application interface after lockdown.}
  \label{fig:Crowd-flow application interface after lockdown}
\BBB
\end{figure}

\medskip
\noindent
\textbf{\textsc{Quest} Applications.} On the secured data, \textsc{Quest} supports the following diverse applications, which monitor/mitigate the spread of COVID-19 (Table~\ref{tab:queries} lists the application in SQL):
\begin{enumerate}[leftmargin=0.01in]
\item 
\textbf{\emph{Location tracing}}: traces all locations that were visited by an infected person in the past 14 days (the possible incubation time of coronavirus). Once the information of an infected person is provided to trapdoor generator, it, first, confirms from the publisher about the infected person, and then, generates trapdoors to find the locations visited by the person during the desired time interval.

\item 
\textbf{\emph{User tracing}:} traces all users that were in the vicinity of an infected person in the past 14 days. Note that this is a natural extension of the previous application, by tracing all people who were at the infected locations at the (\emph{bounded}) interval time (\textit{e}.\textit{g}., +/-15minutes), when an infected person was there.

\item 
\textbf{\emph{Social distancing}:} finds the locations and/or users in the campus that are not following social distancing rule. The idea is to use WiFi dataset to create a predefined occupancy knowledge at the granularity of buildings, floors, and regions within buildings. Now, the dynamic occupancy levels of such buildings (along with the knowledge of the capacity of rooms/floors/buildings) help in establishing to what degree different parts of the buildings are (or have been) occupied. Such a measure can help develop a quantitative metric, a social distance adherence index (\textit{e}.\textit{g}., at UCI, 6 feet distancing requirement was translated roughly into 12.5\% occupancy).

Figure~\ref{fig:Social distancing application interface before lockdown.} shows the interface of social distancing application at UCI before the lockdown was announced. Figure~\ref{fig:Social distancing application interface before lockdown.} shows social distancing at different granularity, such as regions, floors, and buildings, where red-colored dots show that the buildings are not following social distancing rule. Figure~\ref{fig:Social distancing application interface after lockdown.} shows the interface of social distancing application at UCI after the lockdown was announced, where green-colored dots show that the buildings are following social distancing rule.

\item 
\textbf{\emph{Crowd-flow}:} finds locations that were visited by many people in a day, and hence, need frequent cleaning. Note that this is a natural extension of social distancing application. This application provides information to individuals about the number of people visiting a given region over a given time period. Such information can be vital for people wishing to avoid crowded areas and also for the cleaning staff to determine places where disinfecting might be necessary.
Figure~\ref{fig:Crowd-flow application interface after lockdown} shows the interface of crowd-flow application after the lockdown.

\item 
\textbf{\emph{Notification}:} enables all (desired) users to receive notifications, if they are tentative suspects. Note that often when connecting to a WiFi network, it may ask email address or phone number; \textsc{Quest} exploits such information for notifications, (if allowed by the user). 
\end{enumerate}

\section{\textsc{\large \textbf{cQuest}} Protocol}
\label{sec:Computationally-Secure Solution}
This section presents computationally-secure methods, \textsc{cQuest} to encrypt WiFi data and to execute queries on encrypted WiFi data.


\medskip
\noindent\textbf{Key generation.} \textsc{Quest} encrypter generates a symmetric key: $(s_q \oplus \mathit{k}_{\mathit{pko}})||\mathit{attribute}_i$, \textit{i}.\textit{e}, the key is generated for each attribute of $\mathfrak{R}$ by XORing the secret-key of \textsc{Quest} ($s_q$) and public key of organization ($\mathit{k}_{\mathit{pko}}$), and then concatenating with the attribute-id. We denote the key for an attribute $i$ by $k_i$ in Algorithm~\ref{alg:Data Encryption Algorithm}, and unless not clear, we drop the notation $k_i$ from the description.

\subsection*{Data Encryption Method}
\B
Algorithm~\ref{alg:Data Encryption Algorithm} provides pseudocode of proposed data encryption method that is executed at \textsc{Quest} encrypter. It takes tuples of an epoch, produces an encrypted relation $\mathfrak{R}$ with five attributes. Table~\ref{tab:Encrypted WiFi relation for an epoch} shows an example of the produced outputs by Algorithm~\ref{alg:Data Encryption Algorithm}, which works as follows:

\smallskip
\noindent\underline{\emph{Selecting epoch.}} We use the bulk encryption method. Note that WiFi access-points capture time in milliseconds and ping the same device after a certain interval, during which the device can move. These two characteristics of WiFi data capture makes it hard to track a person based on time.\footnote{{\scriptsize For example, a query to find a device's location at 11:00am, cannot be executed in a secure domain, due to unawareness of millisecond-level time generated by APs.}} Thus, we discretize time into equal-length intervals, called \emph{epoch}, and store a special identifier for each interval (that maps to the wall-clock time). An epoch $x$ is denoted by $\Delta_x$ and is identified as a range of begin and end time. All sensor readings during that time period are said to belong to that epoch. There are no gaps between epochs, \textit{i}.\textit{e}., the end time of the previous epoch is the same as the begin time of the next epoch. For simplicity, we identify each epoch by its beginning.


\smallskip
\noindent\underline{\emph{Encrypting device-ids: Attribute $A_{\mathit{id}}$ (Lines~\ref{ln:device_is_not_seen}-\ref{ln:device_is_seen2}).}}
Since a device $d_i$ can appear multiple times in an epoch, we need to prevent the frequency-count attack, while data-at-rest. Also, during query execution, we want to know whether $d_i$ is present in the desired epoch at least once or not. To do so, we encrypt the first appearance of $d_i$ in the epoch as $\mathcal{E}(d_i,1,x)$ and maintain a hash table with value one for $d_i$ in the epoch. Otherwise, we use any random number $r$ and encrypts as $\mathcal{E}(d_i,r)$. We add the epoch-id with $\mathcal{E}(d_i,1,x)$ to make $d_i$'s first appearance indistinguishable in other epochs.

\DontPrintSemicolon
\LinesNotNumbered
\begin{algorithm}[!t]
\scriptsize
\textbf{Inputs:} $\Delta$: duration. $\langle d_i,l_j,t_k\rangle$: A tuple. $\mathcal{H}$: Hash function. $\mathcal{E}$: encryption function.
$\mathrm{PRF}$: a pseudo-random generator.

\textbf{Output:} $\mathfrak{R}(\mathit{A}_{\mathit{id}},\mathit{A}_u, \mathit{A}_L, A_{\mathit{CL}}, \mathit{A}_{\Delta})$: An encrypted relation $\mathfrak{R}$ with five attributes.

\textbf{Variable:} $c_{l_i}$: A counter variable for location $l_i$.

\nl{\bf Function $\mathit{encrypt}(\Delta_x)$}\nllabel{ln:function_create_bucket_workload}
\Begin{
\nl $\forall t_y= \langle d_i,l_j,t_k\rangle \in \Delta_x$: $\ell_i \leftarrow  \mathit{create\_list\_device\_location(\mathrm{distinct}(d_i))}$ \nllabel{ln:crate_list}

\nl $\mathit{HTab}_{\mathit{id}} \leftarrow \mathit{init\_hash\_table\_device}()$,  \nllabel{ln:new_hash_table_device}  $\mathit{HTab}_{L} \leftarrow \mathit{init\_hash\_table\_location}()$  \nllabel{ln:new_hash_table_location}

\nl  \For{$t_y= \langle d_i,l_j,t_k\rangle \in \Delta_x$}{

\nl $r\leftarrow \mathrm{PRF}(\mathit{})$

\nl\lIf{$\mathit{HTab}_{\mathit{id}}[\mathcal{H}(d_i)] \neq 1$}{
    $\mathfrak{R}.\mathit{A}_{\mathit{id}}[y]  \leftarrow  {\mathcal{E}_{k1}(d_i,1,x)}$,
    $\mathfrak{R}.A_u[y]\leftarrow {\mathcal{E}_{k2}(1,y,\Delta_x)}$,
    $\alpha_i[] \leftarrow l_j$ \nllabel{ln:device_is_not_seen}}

\nl \lElseIf{$\mathit{HTab}_{\mathit{id}}[\mathcal{H}(d_i)] == 1 \wedge l_j\notin \alpha_i[]$}{
    $\mathfrak{R}.\mathit{A}_{\mathit{id}}[y] \leftarrow  {\mathcal{E}_{k1}(d_i,r)}$,
    $\mathfrak{R}.A_u[y]\leftarrow {\mathcal{E}_{k2}(1,y,\Delta_x)}$,
    $\alpha_i[] \leftarrow l_j$
    \nllabel{ln:device_is_seen1}}

\nl \lElseIf{$\mathit{HTab}_{\mathit{id}}[\mathcal{H}(d_i)] == 1 \wedge l_j\in \alpha_i[]$}{
    $\mathfrak{R}.\mathit{A}_{\mathit{id}}[y] \leftarrow  {\mathcal{E}_{k1}(d_i,r)}$,
    $\mathfrak{R}.A_u[y]\leftarrow {\mathcal{E}_{k2}(0,r)} $
    \nllabel{ln:device_is_seen2}}

\nl \lIf{$\mathcal{H}(l_j) \notin \mathit{HTab}_{L}  \wedge \mathit{HTab}_{id}[\mathcal{H}(d_i)] \neq 1$}{$\mathit{HTab}_{\mathit{id}}[\mathcal{H}(d_i)] \leftarrow 1$, $c_{l_j}\leftarrow 1$,
    $\mathfrak{R}.\mathit{A}_L[y] \leftarrow  \mathcal{E}_{k3}(l_j,c_{l_j})$,
    $\mathfrak{R}.\mathit{A}_\mathit{CL}[y] \leftarrow  \mathcal{E}_{k4}(r,\ell_i)$ \nllabel{ln:case1}
    }

     \nl \lElseIf{$\mathcal{H}(l_j) \notin \mathit{HTab}_{L}  \wedge \mathit{HTab}_{id}[\mathcal{H}(d_i)] == 1$}{
    $c_{l_j}\leftarrow 1$,
    $\mathfrak{R}.\mathit{A}_L[y] \leftarrow  \mathcal{E}_{k3}(l_j,c_{l_j})$,
    $\mathfrak{R}.\mathit{A}_\mathit{CL}[y] \leftarrow  \mathcal{E}_{k4}(\mathrm{Fake},r)$
     \nllabel{ln:case2}
    }

     \nl \lElseIf{$\mathcal{H}(l_j) \in \mathit{HTab}_{L}  \wedge \mathit{HTab}_{id}[\mathcal{H}(d_i)] \neq 1$}{$\mathit{HTab}_{\mathit{id}}[\mathcal{H}(d_i)] \leftarrow 1$,
    $\mathfrak{R}.\mathit{A}_L[y] \leftarrow \mathcal{E}_{k3}(l_j,c_{l_j}+1)$,
    $\mathfrak{R}.\mathit{A}_\mathit{CL}[y] \leftarrow \mathcal{E}_{k4}(r,\ell_i)$ \nllabel{ln:case3}
    }

    \nl \lElseIf{$\mathcal{H}(l_j) \in \mathit{HTab}_{L}  \wedge \mathit{HTab}_{id}[\mathcal{H}(d_i)] == 1$}{
    $\mathfrak{R}.\mathit{A}_L[y] \leftarrow \mathcal{E}_{k3}(l_j,c_{l_j}+1)$,
    $\mathfrak{R}.\mathit{A}_\mathit{CL}[y] \leftarrow \mathcal{E}_{k4}(\mathrm{Fake},r)$
    \nllabel{ln:case4} 
    }

\nl $\mathfrak{R}.\mathit{A}_{\Delta}[y] \leftarrow  \mathcal{E}_{k5}(\Delta_x)$ \nllabel{ln:encrypt_delta_time}

\nl  $c_{\mathit{max}} \leftarrow \mathit{max}(c_{\mathit{max}}, c_{l_j}), \forall {l_j}$ \nllabel{ln:max_counter_cquest}

}

\nl Delete all hash tables for $\Delta_x$
}
\caption{Data Encryption Algorithm.}
\label{alg:Data Encryption Algorithm}
\end{algorithm}
\setlength{\textfloatsep}{0pt}


\smallskip
\noindent\underline{\emph{Uniqueness of the device: Attribute $A_{\mathit{u}}$ (Lines~\ref{ln:device_is_not_seen}-\ref{ln:device_is_seen2}).}}
To execute applications such as social distancing and crowd-flow, we need to know unique devices at each location in  $\Delta_x$. Thus, when a device $d_i$ appears for the first time at a location in $y^{\mathit{th}}$ tuple, we add its uniqueness by $\mathcal{E}(1,y,\Delta_x)$. (It will avoid \textsc{Quest} to decrypt all encrypted device-ids for knowing distinct devices in $\Delta_x$.)

\smallskip
\noindent\underline{\emph{Encrypting locations: Attributes $A_{\mathit{L}}$ and $A_{\mathit{CL}}$ (Lines~\ref{ln:case1}-\ref{ln:case4}).}} First, we need to produce different ciphertexts for multiple appearances of a  location to prevent frequency-count attack, while data-at-rest. To do so, we use a counter variable for each location and increment by 1, when the same location appears again in a tuple of $\Delta_x$ (and could, also, add $x$, like $d_i$'s encryption). Second, we need to deal with $d_i$ that moves to different locations in $\Delta_x$. Note that based on $E(d_i,1,x)$, we can search only the first appeared location of $d_i$ in $\Delta_x$. Thus, we collect all locations visited by $d_i$ in $\Delta_x$ and add to the combined-locations attribute $A_{\mathit{CL}}$ in a tuple having $E(d_i,1,x)$. We pad the remaining values of $A_{\mathit{CL}}$ by encrypted fake values.

\smallskip
\noindent\underline{\emph{Epoch-ids: Attribute $A_{{\Delta}}$ (Lines~\ref{ln:encrypt_delta_time}).}} Finally, we allocate an identical epoch identifier\footnote{{\scriptsize One may assign the begin time of each epoch as an identifier, \textit{e}.\textit{g}. 4:00, 4:15, and 4:30, while the epoch duration is 15 minutes, or an increasing counter.}} to all tuples belonging to $\Delta_x$ and encrypts it. It allows search based on time, \textit{e}.\textit{g}., based on epoch-id.\footnote{{\scriptsize Based on epoch-ids, we can  execute query to find device's location at any desired time, \textit{e}.\textit{g}., 11:00am.}}


\DontPrintSemicolon
\LinesNotNumbered
\begin{algorithm}[!t]
\scriptsize
\textbf{Inputs:} 
$\mathcal{H}$: Hash function. $\mathcal{E}$: encryption function. $\mathrm{capacity}_{l_i}$: The capacity of location $l_i$. $\mathrm{distanceIndex}$: Maximum \% of allowed people. $\mathrm{Registry}[]$: The list of users allowed sending them notifications.

\textbf{Output:} Answers to queries. 


\nl{\bf Function $\mathit{Location\_Trace}(q(d_i,\mathrm{Time}))$}\nllabel{ln:function_location_trace}
\Begin{

\nl \If{\textbf{\textnormal{Q}} $\leftrightarrow \boldsymbol{\mathcal{P}}$: \textnormal{Verify} $d_i$ \textnormal{Successful} \nllabel{ln:verify}}{


\nl \textbf{Q}: Generate trapdoors $\mathcal{E}(d_i,1,\Delta_t)$: $t$ covers the requested $\mathrm{Time}$ \nllabel{ln:generate_trapdoor}

\nl \textbf{S} $\rightarrow$ \textbf{Q}:
$\mathit{loc}[]\leftarrow$ Location values from $A_{\mathit{CL}}$ corresponding to $\mathcal{E}(d_i,1,\Delta_t)$ \nllabel{ln:fetch_row_using_index_location_encryption}

\nl \textbf{Q}: Decrypt $\mathit{loc}[]$ and produce answers \nllabel{ln:location_decrypt}
}}

\nl{\bf Function $\mathit{User\_Trace}(q(d_i,\mathrm{Time}))$}\nllabel{ln:function_user_trace}
\Begin{
\nl \textbf{Q}: $\mathit{loc}[]\leftarrow \mathit{Location\_Trace}(q(d_i,\mathrm{Time}))$ \nllabel{ln:call_location_trace}

\nl \textbf{Q}: Generate trapdoors: $\forall l_i\in \mathit{loc}$: $\mathcal{E}(l_i,m)$, $m\in \{1, \textnormal{max counter for any location}\}$ \nllabel{ln:generate_trapdoor_user}

\nl \textbf{S} $\rightarrow$ \textbf{Q}:
$\mathit{id}[]\leftarrow$ Values from $A_{\mathit{id}}$ corresponding to $\mathcal{E}(l_i,m)$ \nllabel{ln:s_finds_impacted_user_cquest}

\nl \textbf{Q}: Decrypt $\mathit{id}[]$ and $\mathit{Notification}(\mathit{id}[])$ \nllabel{ln:q_decrypt_know_impacted_users}
}

\nl{\bf Function $\mathit{Social\_Distance}(q(\mathrm{Time}))$}\nllabel{ln:function_social_distance}
\Begin{


\nl \textbf{Q}: Generate trapdoors: $\mathcal{E}(1,y,\Delta_t)$, $y$ is max rows in any epoch, $t$ covers the requested $\mathrm{Time}$ \nllabel{ln:generate_trapdoor_distance}

\nl \textbf{S} $\rightarrow$ \textbf{Q}: $\mathit{loc}[]\leftarrow$ Location values from $A_L$ corresponding to $\mathcal{E}(1,y,\Delta_t)$ \nllabel{ln:join_operation}

\nl \textbf{Q}: $ \forall l_i \in $ Decrypt$(\mathit{loc}[])$, $\mathit{count}_{l_i} \leftarrow  \mathit{count}_{l_i} + 1$ \nllabel{ln:group-by-encryption}

\nl \textbf{Q}: \textbf{if} $\mathit{count}_{l_i} > \mathrm{capacity}_{l_i} \times \mathrm{distanceIndex}$ \textbf{then} Issue alarm \nllabel{ln:issue-alarm-encryption}

}

\nl{\bf Function $\mathit{Crowd\_Flow}(q(\mathrm{Time}))$}\nllabel{ln:function_crowd_flow}
\Begin{

\nl \textbf{Q}: $\mathit{Social\_Distance}(q(\mathrm{Time}))$
}

\nl{\bf Function $\mathit{Notification}(\mathit{id}[])$}\nllabel{ln:function_notification}
\Begin{

\nl \textbf{Q}: \textbf{if} $\forall i, \mathit{id}[i]\in \mathrm{Registry}[]$ \textbf{them} Send notification to $\mathit{id}[i]$ \nllabel{ln:send_notification}
}

\caption{\textsc{cQuest} query execution algorithm.}
\label{alg:Query Execution Algorithm}
\end{algorithm}
\setlength{\textfloatsep}{0pt}

\begin{table*}[!t]
\begin{center}
  \begin{minipage}[t]{.3\linewidth}
  \centering
\begin{tabular}{|l|l|l|l|l|l|l|l|}\hline
&  \texttt{Dev} & \texttt{Loc} & \texttt{Time} \\\hline
1 & $d_1$  & $l_1$ & $t_1$  \\\hline
2 & $d_2$  & $l_2$ & $t_2$  \\\hline
3 & $d_1$  & $l_2$ & $t_1$  \\\hline
4 & $d_1$  & $l_1$ & $t_3$  \\\hline
\end{tabular}
\subcaption{WiFi dataset.}
\label{tab:wif_dataset}
  \end{minipage}
\begin{minipage}[t]{.68\linewidth}
  \centering
\begin{tabular}{|l|l|l|l|l|l|l|}\hline
 & $A_{\mathit{id}}$            & $A_{u}$                  & $A_{L}$                   & $A_{\mathit{CL}}$              & $A_{\Delta}$ \\\hline

1 & $\mathcal{E}_{k1}(d_1,1,x)$ & $\mathcal{E}_{k2}(1,1,x)$ & $\mathcal{E}_{k3}(l_1,1)$ & $\mathcal{E}_{k4}(r,l_1,l_2)$   & $\mathcal{E}_{k5}(x)$  \\\hline

2 & $\mathcal{E}_{k1}(d_2,1,x)$ & $\mathcal{E}_{k2}(1,2,x)$ & $\mathcal{E}_{k3}(l_2,1)$ & $\mathcal{E}_{k4}(r,l_1)$       & $\mathcal{E}_{k5}(x)$ \\\hline

3 & $\mathcal{E}_{k1}(d_1,r,x)$ & $\mathcal{E}_{k2}(1,3,x)$ & $\mathcal{E}_{k3}(l_2,2)$ & $\mathcal{E}_{k4}(\mathrm{Fake},3)$ & $\mathcal{E}_{k5}(x)$ \\\hline

4 & $\mathcal{E}_{k1}(d_1,r,x)$ & $\mathcal{E}_{k2}(0,r)$   & $\mathcal{E}_{k3}(l_1,2)$ & $\mathcal{E}_{k4}(\mathrm{Fake},4)$ & $\mathcal{E}_{k5}(x)$ \\\hline
  \end{tabular}
  \subcaption{Encrypted WiFi relation for an epoch.}
  \label{tab:Encrypted WiFi relation for an epoch}
    \end{minipage}

\begin{minipage}[t]{.35\linewidth}
  \centering
\begin{tabular}{|l|l|l|l|l|l|l|}\hline
 & $A_{\mathit{smid}}$  & $A_{\mathit{sid}}$ & $A_{\mathit{su}}$         & $A_{\mathit{smL}}$          & $A_{\mathit{sL}}$                          & $A_{\Delta}$ \\\hline

1 & $\mathrm{SSS}(d_1)$ & $\mathrm{S}(d_1)$  & $\mathrm{S}(1)$ & $\mathrm{SSS}(l_1)$ & $\mathrm{S}(l_1)$ & $x$  \\\hline

2 & $\mathrm{SSS}(d_2)$ & $\mathrm{S}(d_2)$  & $\mathrm{S}(1)$ & $\mathrm{SSS}(l_2)$ & $\mathrm{S}(l_2)$ & $x$ \\\hline

3 & $\mathrm{SSS}(d_1)$ & $\mathrm{S}(d_1)$  & $\mathrm{S}(1)$ &  $\mathrm{SSS}(l_2)$ & $\mathrm{S}(l_2)$ & $x$ \\\hline

4 & $\mathrm{SSS}(d_1)$ & $\mathrm{S}(d_1)$  & $\mathrm{S}(0)$ &  $\mathrm{SSS}(l_1)$ & $\mathrm{S}(l_1)$ & $x$ \\\hline
  \end{tabular}
  \subcaption{Secret-shared WiFi relation for an epoch.}
  \label{tab:Secret-shared WiFi relation for an epoch}
    \end{minipage}
\end{center}
\caption{Original WiFi dataset, encrypted WiFi dataset using Algorithm~\ref{alg:Data Encryption Algorithm}, and secret-shared WiFi dataset using Algorithm~\ref{alg:Secret-share creation algorithm}.}
\label{table:Original WiFi dataset and the encrypted WiFi dataset using Algorithm1}
\BB
\end{table*}

\subsection*{Query Execution}
\B
Table~\ref{tab:queries} shows applications supported by \textsc{Quest} in SQL, and Algorithm~\ref{alg:Query Execution Algorithm} explains trapdoor generation process at \textsc{Quest} (denoted by \textbf{Q}) for those applications and query execution at the public server (denoted by \textbf{S}).\footnote{{\scriptsize For simplicity,  we denote a queried device-id by $d_i$. In practice, depending upon the publisher $\mathcal{P}$, such a device-id might be encrypted. in which case \textbf{Q} may need to securely obtain the real device-id from $\mathcal{P}$  during verification (line~\ref{ln:verify} Algorithm~\ref{alg:Query Execution Algorithm})}}

\medskip
\noindent\textbf{{\emph{Location tracing} (lines~\ref{ln:function_location_trace_ss}-\ref{ln:user_get_location_ss}).}} First, \textbf{Q} verifies the identity of the infected user device $d_i$ from the publisher $\mathcal{P}$ (line~\ref{ln:verify}). Then, \textbf{Q} creates and sends trapdoors for $d_i$ as: $\mathcal{E}(d_i,1,\Delta_t)$, where $t$ is the epoch-identifiers that can cover the desired queried time (line~\ref{ln:generate_trapdoor}). \textbf{S} executes a selection query for fetching the values of $A_{\mathit{CL}}$ column corresponding to all encrypted query trapdoors (line~\ref{ln:fetch_row_using_index_location_encryption}). The answers to the selection query are given to \textbf{Q} that decrypts them to know the impacted locations (line~\ref{ln:location_decrypt}).

\noindent
\textit{\underline{Example~\ref{sec:Computationally-Secure Solution}.1.}} Suppose $d_1$ belongs to an infected person in Table~\ref{tab:Encrypted WiFi relation for an epoch}, and all four tuples belong to an identical epoch $x$. To execute location tracing, \textbf{Q} creates trapdoor for $d_1$, as: $\mathcal{E}(d_1,1,x)$. \textbf{S} checks the trapdoor in $A_{\mathit{id}}$ column and sends the corresponding value of $A_{\mathit{CL}}$ column, \textit{i}.\textit{e}., $\mathcal{E}_{k4}(r, l_1,l_2)$ to \textbf{Q}. On decrypting the received answer, \textbf{Q} knows the impacted locations as $l_1$ and $l_2$.

\medskip
\noindent\textbf{{\emph{User tracing} (lines~\ref{ln:function_user_trace}-\ref{ln:q_decrypt_know_impacted_users}).}} First, \textbf{Q} executes $\mathit{Location\_Trace}()$ to know the impacted locations by the infected person (line~\ref{ln:call_location_trace}). Then, \textbf{Q} creates trapdoors for all such locations (line~\ref{ln:generate_trapdoor_user}), as: $\mathcal{E}(l_i,m)$, where $l_i$ is the $i^{\mathit{th}}$ impacted location and $m$ is the maximum counter value for any location in any epoch, as obtained in Algorithm~\ref{alg:Data Encryption Algorithm}'s line~\ref{ln:max_counter_cquest}.\footnote{{\scriptsize Generating and sending trapdoors for impacted locations equals to the maximum counter value may incur computation and communication overheads. Thus, we will suggest an optimization for preventing this.}}  \textbf{S} executes a selection query for the trapdoor (or a join query between a table having all trapdoors and another table having the encrypted WiFi data) to know the corresponding values of $A_{\mathit{id}}$ column (line~\ref{ln:s_finds_impacted_user_cquest}). All such values are transmitted to \textbf{Q} that decrypts them to know the final answer (line~\ref{ln:q_decrypt_know_impacted_users}). If any of the impacted users have subscribed to notification service, then they are informed.

\noindent
\textit{\underline{Example~\ref{sec:Computationally-Secure Solution}.2.}} Suppose, we wish to know the impacted people that may in contact with the infected person whose device-id is $d_1$. From Example~\ref{sec:Computationally-Secure Solution}.1, we know that $\langle l_1,l_2\rangle$ are the impacted locations. Suppose the maximum counter value for any location ($c_\mathit{max}$) is two. Thus, \textbf{Q} generates trapdoors as follows: $\mathcal{E}(l_1,1)$, $\mathcal{E}(l_1,2)$, $\mathcal{E}(l_2,1)$, $\mathcal{E}(l_2,2)$, and sends them to \textbf{S}. \textbf{S} executes a selection query over $A_{L}$ column for such trapdoors and sends device-ids from $A_{\mathit{id}}$ column, corresponding to the trapdoors. After the decryption, \textbf{Q} knows that $d_2$ is the device of a person that was in contact with the infected person whose device-id is $d_1$.

\medskip
\noindent\textbf{{\emph{Social distancing} (lines~\ref{ln:function_social_distance}-\ref{ln:issue-alarm-encryption}).}} \textbf{Q} generates and sends trapdoors $\mathcal{E}(1,y,\Delta_t)$ to \textbf{S} to find the unique devices in the desired epochs (line~\ref{ln:generate_trapdoor_distance}).\footnote{{\scriptsize Sending trapdoors that are equal to the number of tuples in the desired epoch may incur communication overheads. Soon, we will provide optimizations for avoiding such trapdoor generation and transmission.}} \textbf{S} executes a selection query for the trapdoors (or a join query between a table having all trapdoors and another table having the encrypted WiFi data) to know the unique devices in the desired epochs and sends the qualified values from $A_{L}$ column (line~\ref{ln:join_operation}) to \textbf{Q}. \textbf{Q}  decrypts the received locations and counts the appearance of each location (line~\ref{ln:group-by-encryption}). Then, \textbf{Q} issues an alarm, if the counter value for a location exceeds the predefined rule for social distancing, denoted by $\mathrm{distanceIndex}$ (line~\ref{ln:issue-alarm-encryption}).

\textit{Aside.} Note that we can also know the devices that do not follow the predefined rule for social distancing, by fetching the qualified values from $A_{\mathit{id}}$ along with values of $A_L$.

\noindent
\textit{\underline{Example~\ref{sec:Computationally-Secure Solution}.3.}} Assume that if more than one person appear at a location during a given epoch, then it shows that people at the location are not following the  predefined social distancing rules, \textit{i}.\textit{e}.,
in this example, $\mathrm{distanceIndex}=1$.
\textbf{Q} generates the following four trapdoors: $\mathcal{E}(1,1,x)$, $\mathcal{E}(1,2,x)$, $\mathcal{E}(1,3,x)$, and $\mathcal{E}(1,4,x)$. Based on these trapdoors, \textbf{S} sends $\mathcal{E}(l_1,1)$, $\mathcal{E}(l_2,1)$, and $\mathcal{E}(l_2,2)$. On receiving the encrypted location values, \textbf{Q} decrypts them, counts the number of each location, and finds that the location $l_2$ is not following the social distancing rule.

\medskip
\noindent
\textbf{Information leakage discussion.} Although the data-at-rest does not reveal any information, the query execution reveals access-patterns (like SSEs or SGX-based systems~\cite{DBLP:conf/eurosec/GotzfriedESM17,DBLP:conf/ccs/WangCPZWBTG17,opaque,DBLP:journals/pvldb/EskandarianZ19,DBLP:journals/corr/abs-2002-05097}). Thus, an adversary, by just observing the query execution, may learn additional information, \textit{e}.\textit{g}.,
which of the tuples correspond to an infected person (by observing $\mathit{Location\_Trace}$), how many people may get infected by an infected person (by observing $\mathit{User\_Trace}$), which tuples correspond to unique devices by observing queries on $A_u$ or $A_{\mathit{CL}}$, and which locations are frequently visited by users (by observing $\mathit{Social\_Distance}$).
Also, since \textsc{cQuest} is based on encryption, a computationally-efficient adversary can break the underlying encryption technique.

\medskip
\noindent\textbf{Pros.} Though the approach is simple, \textsc{cQuest} maintains hash tables during encryption of tuples belonging to an epoch. Nevertheless, the size of hash tables is small for an epoch, (see \S\ref{sec:Experimental Evaluation}). \textsc{cQuest} efficiently deals with dynamic data, due to independence from an explicit indexable data structure, (unlike indexable SSE techniques~\cite{DBLP:journals/pvldb/LiLWB14,DBLP:conf/icde/LiL17} that require to rebuild the entire index due to data insertion at the trusted size). Algorithm~\ref{alg:Query Execution Algorithm} avoids reading, decrypting the entire data of an epoch to execute a query, (unlike SGX-based systems~\cite{opaque}); thus, saves computational overheads. Also, the key generation by XORing $s_q$ and $k_{\mathit{pko}}$ prevents the adversary to learn any information by observing at the encrypted data belonging to two different organizations, since one of the keys will be surely different at different organizations. 

\medskip
\noindent
\textbf{Cons.} Algorithm~\ref{alg:Data Encryption Algorithm} increases the dataset size by adding two additional columns.
Algorithm~\ref{alg:Query Execution Algorithm} reveals access-patterns; hence, the adversary may deduce information based on access-patterns. Similar to DaS model~\cite{DBLP:conf/icde/HacigumusMI02}, the trapdoor generator has to decrypt the retrieved tuples, possibly to filter them, and to execute a small computation (\textit{e}.\textit{g}., group by operation line~\ref{ln:group-by-encryption} of Algorithm~\ref{alg:Query Execution Algorithm}). Also, as we mentioned earlier that \textsc{Quest} has a limitation that encrypter or trapdoor modules should not be tampered, by anyone, likewise DaS model~\cite{DBLP:conf/icde/HacigumusMI02}.

\medskip
\noindent
\textbf{Optimizations.} We provide four optimizations: two for trapdoor generation in $\mathit{User\_Trace}()$, and the other two for trapdoor generation for $\mathit{Social\_Distance}()$. \S\ref{sec:Experimental Evaluation} will show the impact of such optimizations.

\smallskip
\noindent
\textbf{\textit{Location counters.}} Note that Line~\ref{ln:generate_trapdoor_user} of Algorithm~\ref{alg:Query Execution Algorithm} requires us to generate the number of trapdoors equals to the maximum counter values (\textit{i}.\textit{e}., maximum connection events at a location in any epoch (Line~\ref{ln:max_counter_cquest} of Algorithm~\ref{alg:Data Encryption Algorithm})). It may incur overhead in generating trapdoors and sending them to the server. Thus, we can reduce the number of trapdoors by keeping two types of counters: (\textit{i}) \emph{counter per epoch} to contain the maximum connection events at a location in each epoch, and (\textit{ii}) \emph{counter per epoch and per location} to contain the maximum connection events at each location in each epoch.

\smallskip
\noindent
\textbf{\textit{Trapdoor generation for uniqueness finding.}} Line~\ref{ln:generate_trapdoor_distance} Algorithm~\ref{alg:Query Execution Algorithm} requires \textsc{Quest}'s encrypter to generate and send the number of trapdoors equals to the maximum number of tuples in any epoch. We can avoid sending many trapdoors by encrypting uniqueness of the device, as follows: $\mathcal{E}_k(\mathcal{E}_{k^{\prime}}(1,\Delta_x),y)$ (at Line~\ref{ln:device_is_not_seen}-\ref{ln:device_is_seen2} of Algorithm~\ref{alg:Data Encryption Algorithm}), where $k$ is known to \textbf{S} and ${k^{\prime}}=(s_q\oplus k_{\mathit{pko}})||\mathit{attribute}$ is unknown to \textbf{S}. Thus, for social distancing query execution, \textbf{Q} needs to send to \textbf{S} only $\gamma=\mathcal{E}_{k^{\prime}}(1,\Delta_x)$, and then, \textbf{S} can generate all the desired trapdoors as $\mathcal{E}_k(\gamma,y)$, where $y$ is the number of rows in the desired epoch.

In the above-mentioned optimization, \textsc{Quest}'s encrypter does not need to generate all trapdoors and sends them the server, and the server will generate the desired trapdoors. While it will reduce the communication cost, the computation cost at the server will remain identical to the method given in Algorithm~\ref{alg:Query Execution Algorithm} Lines~\ref{ln:s_finds_impacted_user_cquest}. Thus, in order to reduce the computation cost at the server, we can also outsource the hash table created for locations ($\mathit{HTab}_L$, Line~\ref{ln:new_hash_table_location} Algorithm~\ref{alg:Data Encryption Algorithm}), after each epoch. Now, to execute the social distancing application, \textsc{Quest} needs to ask the server to send the encrypted hash tables for all the desired epochs. Since the hash table contains the number of unique devices at each location, it will provide the correct answer to the social distancing application after decryption at \textsc{Quest}.

\section{\textsc{\large\textbf{iQuest}} Protocol}
\label{sec:Information-Theoretically Secure Solution}
To overcome the information leakages due to \textsc{cQuest}, we provide a completely secure solution, \textsc{iQuest} that is based on string-matching operation~\cite{DBLP:conf/ccs/DolevGL15} on secret-shares~\cite{DBLP:journals/cacm/Shamir79}.

\medskip
\noindent
\textbf{Background: String-matching over secret-shares.} As a building block, first, we explain the string matching of Dolev et al.~\cite{DBLP:conf/ccs/DolevGL15} using the following example.

\noindent\textit{Data Owner: outsourcing searchable-secret-share (SSS).} Assume there are only two symbols: X and Y; thus, X and Y can be written as $\langle 1,0\rangle$ and $\langle 0,1\rangle$. Suppose, the owner wishes to outsource Y; thus, she creates \emph{unary vector} $\langle 0,1\rangle$. But, to hide exact numbers in $\langle0,1\rangle$, she creates secret-shares of each number using polynomials of an identical degree (see Table~\ref{tab:Secret-shares by DB owner}) and sends shares to servers.

\bgroup
\def\arraystretch{.86}
\begin{table}[h]
\centering
  \scriptsize
\begin{tabular}{|p{.8cm}|l|l|l|l|}\hline

  Values & Polynomials & I$^\mathrm{st}$ shares & II$^\mathrm{nd}$ shares & III$^\mathrm{rd}$ shares \\\hline
  0 & $0+2x$ & 2 &  4 & 6  \\\hline
  1 & $1+8x$ & 9 & 17 & 25  \\\hline
\end{tabular}
\caption{Secret-shares of $\langle 1,0,0,1\rangle$, created by the owner.}
\label{tab:Secret-shares by DB owner}
\end{table}
\egroup

\noindent\textit{User: SSS query generation.} Suppose a user wishes to search for Y. She creates unary vectors of Y as $\langle0,1\rangle$, and then, creates secret-shares of each number of $\langle0,1\rangle$ using any polynomial of the same degree as used by the owner (see Table~\ref{tab:Secret-shares by user}). Note that since a user can use any polynomial, it prevents an adversary to learn an equality condition by observing query predicates and databases. 

\bgroup
\def\arraystretch{.86}
\begin{table}[h]
  \centering
  \scriptsize
\begin{tabular}{|p{.8cm}|l|l|l|l|}\hline

  Values & Polynomials & I$^\mathrm{st}$ shares & II$^\mathrm{nd}$ shares & III$^\mathrm{rd}$ shares \\\hline

  0 & $0+3x$ & 3 & 6 & 9   \\\hline
  1 & $1+7x$ & 8 & 15& 22 \\\hline
 \end{tabular}
\caption{Secret-shares of $\langle 1,0,0,1\rangle$, created by the user.}
\label{tab:Secret-shares by user}
\end{table}
\egroup

\noindent\textit{Servers: String-matching operation.} Now, each server has a secret-shared database and a secret-shared query predicate. For executing the string-matching operation, the server performs bit-wise multiplication and then adds all outputs of multiplication (see Table~\ref{tab:cloud multiply}). 

\bgroup
\def\arraystretch{.86}
\begin{table}[!h]
  \centering
  \scriptsize
\begin{tabular}{|l|l|l|l|l|l|l|l|}\hline
  Server 1             & Server 2           & Server 3 \\\hline
  $2\times 3 =  6$    & $4\times  6 = 24$ & $6\times9=54$ \\\hline

  $9\times 8=72$ & $17\times  15 =255$ &  $25\times22=550$ \\\hline

78 & 279 & 604 \\\hline

\end{tabular}
\caption{Servers' computation.}
\label{tab:cloud multiply}
\end{table}
\egroup

\noindent\textit{User: result reconstruction.} User receives results from all servers and performs Lagrange interpolation~\cite{corless2013graduate} to obtain final answers:

\centerline{\scriptsize
$
\frac{(x-2)(x-3)}{(1-2)(1-3)}\times 72 +
\frac{(x-1)(x-3)}{(2-1)(2-3)}\times 255 +
\frac{(x-1)(x-2)}{(3-1)(3-2)}\times 550  = 1
$}
\noindent Now, if the final answer is 1, it shows that the secret-shared database at the server matches the user query.

\subsection*{Data Outsourcing Method}
\B
\textsc{iQuest} uses Algorithm~\ref{alg:Secret-share creation algorithm} for creating secret-shares of input WiFi relation $R$. \emph{Note that Algorithm~\ref{alg:Secret-share creation algorithm} when creating SSS or Shamir's secret-shares of a value (denoted by $\mathrm{SSS}(v)$ and $\mathrm{S}(v)$, respectively), randomly selects a polynomial of an identical degree.} Table~\ref{tab:Secret-shared WiFi relation for an epoch} shows an example of the output of Algorithm~\ref{alg:Secret-share creation algorithm}. Algorithm~\ref{alg:Secret-share creation algorithm} selects an epoch duration (like \textsc{cQuest} (\S\ref{sec:Computationally-Secure Solution})) and produces an $i^{\mathit{th}}$ secret-shared relation $S(\mathfrak{R})_i$ with six attributes, denoted by  $A_{\mathit{smid}}$, $A_{\mathit{sid}}$, $A_{\mathit{su}}$,
$A_{\mathit{smL}}$, $A_{\mathit{sL}}$, and $A_{\Delta}$.
Note that \emph{if the adversary cannot collude any two non-communicating servers, then we can use polynomials of degree one}, and in this case, there is no need to create more than $2l+2$ shares, where $l$ is the maximum length of a secret, to obtain an answer to a query in one communication round between the user and  servers. 
Algorithm~\ref{alg:Secret-share creation algorithm} works as follows:

\smallskip
\noindent
\underline{\emph{Secret-shares of devices: Attributes $A_{\mathit{smid}}$, $A_{\mathit{sid}}$ (Lines~\ref{ln:find_last_v_bit_sss}-\ref{ln:create_share_device_id}).}}
We create two types of shares of each device id, one is SSS that is used for string matching operation and stored in $A_{\mathit{smid}}$, and another is just a Shamir's secret-share of the entire device-id stored in $A_{\mathit{sid}}$.
The purpose of storing the same device-id in two different formats is to speed-up the computation. Particularly, values in $A_{\mathit{smid}}$ help in string-matching operation, when we want to search for a device-id (\textit{e}.\textit{g}., location tracing application), and values in $A_{\mathit{sid}}$ helps in fetching the device-id (\textit{e}.\textit{g}., user tracking application for retrieving device-ids based on infected locations).

\emph{Aside.} Recall that creating secret-shares for string matching requires to convert the device-id into a \emph{unary vector}; as shown in Table~\ref{tab:Secret-shares by DB owner}. However, it increases the length of device-ids significantly (\textit{i}.\textit{e}., $12 \times 16= 192$, often a device-id (MAC-ID) contains 12 hexadecimal digits (a combination of numbers $0,1,\ldots,9$ and alphabets A, B, $\ldots$ F), and thus, every single MAC-ID digit will use a unary vector of size 16). Thus, we, first, execute a hash function (only known to \textsc{iQuest}) on each device-id to map to a smaller length string, by taking the last $v<12$ digits of the digest. Hashing may result in a collision, by mapping two different device-ids to the same digest, with a very low probability. For example, for a 256-bit hash function, the probability of collision in mapping all possible 32-bit integers is $ 2^{64}/2^{256+1} =  1/2^{193}$, which is negligible.

\smallskip
\noindent
\underline{\emph{Uniqueness of devices: Attribute $A_{\mathit{su}}$ (Lines~\ref{ln:device_is_not_seen_ss}-\ref{ln:device_is_seen_ss2}).}} Similar to \textsc{cQuest}'s Algorithm~\ref{alg:Data Encryption Algorithm}, we assign value one when $d_i$ appears for the first time at a location in an epoch; otherwise, zero. After that we create secret-shares of the value.

\smallskip
\noindent
\underline{\emph{Secret-shares of location: Attributes $A_{\mathit{smid}}$, $A_{\mathit{sid}}$ (Line~\ref{ln:create_share_location_ss}).}} Likewise two types of secret-shares for device-ids, we create two types of shares of each location, one is SSS -- stored in $A_{\mathit{smL}}$, and another is a Shamir's secret-share of the location stored in $A_{\mathit{sL}}$.

\smallskip
\noindent\underline{\emph{Outsourcing epoch-ids: Attributes $A_{\Delta}$ (Line~\ref{ln:delta_time_ss}).}} Finally, for all tuples of $\Delta_x$, we outsource an epoch identifier in cleartext.

\DontPrintSemicolon
\LinesNotNumbered
\begin{algorithm}[!t]
\scriptsize
\textbf{Inputs:} $\Delta$: duration. $\langle d_i,l_j,t_k\rangle$: A tuple. $\mathcal{H}$: A hash function known to only \textsc{iQuest}. 
$z$: a secret of proxy, unknown to organization. 

\textbf{Output:} $S(\mathfrak{R})_i(A_{\mathit{smid}}, A_{\mathit{sid}}, A_{\mathit{su}}, A_{\mathit{smL}}, A_{\mathit{sL}}, A_{\Delta})$: An $i^{\mathit{th}}$ encrypted relation $R$ with six attributes.

\textbf{Functions:} $\mathrm{SSS}(\mathit{v})$: A function for creating searchable secret-shares of $\mathit{v}$. 
$\mathrm{S}(\mathit{v})$: A function for creating Shamir's secret-shares of $\mathit{v}$.

\nl{\bf Function $\mathit{create\_shares}(\Delta_x)$}\nllabel{ln:function_create_bucket_workload}
\Begin{
\nl $\mathit{HTab}_{\mathit{id}} \leftarrow \mathit{init\_hash\_table\_device}()$ \nllabel{ln:new_hash_table_device_device_sss}  

\nl  \For{$t_y= \langle d_i,l_j,t_k\rangle \in \Delta_x$}{

\nl $\mathit{val}\leftarrow \mathit{last\_v\_bits}(\mathcal{H}(d_i))$ \nllabel{ln:find_last_v_bit_sss}

\nl $\mathfrak{R}.A_{\mathit{smid}}[y]\leftarrow \mathrm{SSS}(\mathit{val})$,
    $\mathfrak{R}.A_{\mathit{sid}}[y] \leftarrow   \mathrm{S}(\mathit{val})$ \nllabel{ln:create_share_device_id}

\nl\lIf{$\mathit{HTab}_{\mathit{id}}[\mathcal{H}(d_i)] \neq 1$}{
$\mathfrak{R}.A_{\mathit{su}}[y] \leftarrow  \mathrm{S}({1})$, $\alpha_i[] \leftarrow l_j$
\nllabel{ln:device_is_not_seen_ss}}

\nl \lElseIf{$\mathit{HTab}_{\mathit{id}}[\mathcal{H}(d_i)] == 1 \wedge l_j\notin \alpha_i[]$}{
    $\mathfrak{R}.A_{\mathit{su}}[y] \leftarrow  \mathrm{S}({1})$, $\alpha_i[] \leftarrow l_j$
    \nllabel{ln:device_is_seen_ss1}}

\nl \lElseIf{$\mathit{HTab}_{\mathit{id}}[\mathcal{H}(d_i)] == 1 \wedge l_j\in \alpha_i[]$}{
    $\mathfrak{R}.A_{\mathit{su}}[y] \leftarrow  \mathrm{S}({0})$
    \nllabel{ln:device_is_seen_ss2}}

\nl $\mathfrak{R}.A_{\mathit{smL}}[y]\leftarrow \mathrm{SSS}(\mathit{l_j})$,
    $\mathfrak{R}.A_{\mathit{sL}}[y] \leftarrow   \mathrm{S}(\mathit{l_j})$
    \nllabel{ln:create_share_location_ss}

\nl $\mathfrak{R}.\mathit{A}_{\Delta}[y] \leftarrow  \mathit{identifier}(\Delta_x)$, \nllabel{ln:delta_time_ss} $\mathit{HTab}_{\mathit{id}}[\mathcal{H}(d_i)] \leftarrow 1$

}
\caption{Secret-share creation algorithm.}
\label{alg:Secret-share creation algorithm}
}
\end{algorithm}
\setlength{\textfloatsep}{0pt}

\medskip
\noindent
\textbf{Differences between data outsourcing methods of \textsc{cQuest} and \textsc{iQuest}.}
Though \textsc{cQuest} is an encryption-based method and \textsc{iQuest} is a secret-sharing-based method, they, also, differ the way of keeping metadata (in Algorithms~\ref{alg:Data Encryption Algorithm} and~\ref{alg:Secret-share creation algorithm}).
First, \textsc{iQuest} does not keep a hash table for locations to maintain their occurrences in tuples of an epoch.
Second, \textsc{iQuest} does not need to first find all locations visited by a device during an epoch and adds them in a special attribute; hence, \textsc{iQuest} does not keep attribute $A_{\mathit{CL}}$.
Note that these differences occur, due to exploiting the capabilities of SSS and selecting different polynomials for creating shares of any value, thereby, different occurrences of an identical value appear different in secret-shared form.

\DontPrintSemicolon
\LinesNotNumbered
\begin{algorithm}[!t]
\scriptsize
\textbf{Inputs:} 
$\mathcal{H}$: Hash function. $\mathrm{capacity}_{l_i}$: The capacity of location $l_i$. $\mathrm{distanceIndex}$: Maximum \% of allowed people.

\textbf{Output:} Answers to queries. 

\textbf{Functions:} $\mathrm{SSS}(\mathit{v})$ and $\mathrm{S}(\mathit{v})$: From Algorithm~\ref{alg:Secret-share creation algorithm}. 
$\mathrm{interpolate(shares)}$: An interpolation function that takes shares as inputs and produces the secret value.

\nl{\bf Function $\mathit{Location\_Trace}(q(d_i,\mathrm{Time}))$} \nllabel{ln:function_location_trace_ss}
\Begin{
\nl \textbf{Q} $\leftrightarrow \boldsymbol{\mathcal{P}}$: Verify $d_i$ \nllabel{ln:verify_ss}

\nl \textbf{Q} $\rightarrow$ \textbf{S}: $\gamma \leftarrow \mathrm{SSS}(d_i)$, $\Delta_t$: $t$ covers the requested $\mathrm{Time}$ \nllabel{ln:generate_trapdoor_ss}

\nl \textbf{S}:
$\mathit{sLoc}[] \leftarrow (A_{\mathit{smid}}[j] \otimes \gamma ) \times A_{\mathit{sL}}$,
$j \in \{ 1,y \}$, $y =$ \#tuples in $\Delta_t$ \nllabel{ln:server_trace_location_ss}

\nl \textbf{Q}: $\mathit{location}[]\leftarrow \mathrm{interpolate}(\mathit{sLoc}[])$ \nllabel{ln:user_get_location_ss}
}

\nl{\bf Function $\mathit{User\_Trace}(q(d_i,\mathrm{Time}))$}\nllabel{ln:function_user_trace_ss}
\Begin{
\nl \textbf{Q}: $\mathit{location}[] \leftarrow \mathit{Location\_Trace}(q(d_i,\mathrm{Time}))$ \nllabel{ln:user_call_location_trace_ss}

\nl \textbf{Q} $\rightarrow$ \textbf{S}: $\mathit{sssLoc}[]\leftarrow \mathrm{SSS}(\mathit{location}[])$,  $\Delta_t$: $t$ covers the requested $\mathrm{Time}$ \nllabel{ln:user_convert_share_location_trace_ss}

\nl \textbf{S}: $\forall i\in\{1,|\mathit{sssLoc}[]|\}$,
                $\forall j\in\{1,y\}$, $y =$ \#tuples in $\Delta_t$,
                $\mathit{sID}[i,j]\leftarrow (\mathit{sssloc}[i] \otimes A_{\mathit{smL}}[j])\times A_{\mathit{sid}}[j]$ \nllabel{ln:s_find_impacted_person_ss}

\nl \textbf{Q}: $\mathit{id}[]\leftarrow \mathrm{interpolate}(\mathit{sID}[\ast,\ast])$, $\forall i\in\{1,|\mathit{sID}[\ast,\ast]|\}$ \nllabel{ln:user_get_impacted_person_ss}

\nl \textbf{Q}: $\mathit{Notification}(\mathit{id}[])$ of \textbf{Algorithm~\ref{alg:Query Execution Algorithm}} \nllabel{ln:user_notification_ss}
}

\nl{\bf Function $\mathit{Social\_Distance}(q(\mathrm{Time}))$}\nllabel{ln:function_social_distance_ss}
\Begin{

\nl \textbf{Q} $\rightarrow$ \textbf{S}: $\Delta_t$: $t$ covers the requested $\mathrm{Time}$ \nllabel{ln:quest_sends_delta}

\nl \textbf{S} $\rightarrow$ \textbf{Q}: $\mathit{sLoc}[j]\leftarrow A_{\mathit{su}}[j] \times A_{\mathit{sL}}[j]$, $\forall j \in \Delta_t$ \nllabel{ln:social_server}

\nl \textbf{Q}:  $\mathit{location}[]\leftarrow \mathrm{interpolate}(\mathit{sLoc}[])$ \nllabel{ln:q_interpolate_social_dis_location_ss}

\nl \textbf{Q}: $ \forall l_i \in \mathit{location}[]$, $\mathit{count}_{l_i} \leftarrow  \mathit{count}_{l_i} + 1$ \nllabel{ln:q_groupby_social_dis_ss}

\nl \textbf{Q}: \textbf{if} $\mathit{count}_{l_i} > \mathrm{capacity}_{l_i} \times \mathrm{distanceIndex}$ \textbf{then} Issue alarm \nllabel{ln:social_distance_issue_alarm_ss}
}

\nl{\bf Function $\mathit{Crowd\_Flow}(q(\mathrm{Time}))$}\nllabel{ln:function_crowd_flow_ss}
\Begin{

\nl \textbf{Q}: $\mathit{Social\_Distance}(q(\mathrm{Time}))$

}
\caption{\textsc{iQuest} query execution algorithm.}
\label{alg:iquest_Query Execution Algorithm_sss}
\end{algorithm}
\setlength{\textfloatsep}{0pt}

\subsection*{Query Execution}
Algorithm~\ref{alg:iquest_Query Execution Algorithm_sss} explains secret-shared query generation at \textsc{iQuest} (denoted by \textbf{Q}), query execution at the server (denoted by \textbf{S}), and final processing before producing the answer at \textbf{Q}. Note that in Algorithm~\ref{alg:iquest_Query Execution Algorithm_sss}, \emph{$\otimes$ denotes string-matching operation and $\times$ denotes normal arithmetic multiplication}. Below, we explain query execution for different applications over secret-shares.

\medskip
\noindent\textbf{{\emph{Location tracing} (lines~\ref{ln:function_location_trace_ss}-\ref{ln:user_get_location_ss}).}} First, \textbf{Q} verifies the device id $d_i$ (as the real device of an infected person) from the publisher $\mathcal{P}$ (line~\ref{ln:verify_ss}). Then, \textbf{Q} creates SSS of $d_i$ (denoted by $\gamma$) and sends it to each non-communicating server along with the desired epoch identifier (line~\ref{ln:generate_trapdoor_ss}). Each server executes string-matching operation over each value of $A_{\mathit{smid}}$ against $\gamma$ in the desired epoch, and it will result in either 0 or 1 (recall that string-matching operation results in only 0 or 1 of \emph{secret-shared form}). Then, the $i^{\mathit{th}}$ result of string-matching operation is multiplied by $i^{\mathit{th}}$ value of $A_{\mathit{sL}}$, resulting in the secret-shared location, if impacted by the user; otherwise, the secret-shared location value will become 0 of secret-shared form (line~\ref{ln:server_trace_location_ss}). Finally, \textbf{Q} receives shares from all servers, interpolates them, and it results in all locations visited by the infected person (line~\ref{ln:user_get_location_ss}).

\noindent
\textit{\underline{Example~\ref{sec:Information-Theoretically Secure Solution}.1.}} Suppose $d_1$ belongs to an infected person in Table~\ref{tab:Secret-shared WiFi relation for an epoch}. To execute location tracing, \textbf{Q} generates SSS of $d_1$, say $\gamma$. \textbf{S} checks $\gamma$ against the four shares (via string-matching operation) in $A_{\mathit{smid}}$ and results in $\langle1,0,1,1\rangle$ (of secret-shared form) that is position-wise multiplied by  $\langle\mathrm{S}(l_1),\mathrm{S}(l_2),\mathrm{S}(l_2),\mathrm{S}(l_1)\rangle$. Thus, \textbf{S} sends $\langle l_1,0,l_2,l_1\rangle$ of secret-shared form to \textbf{Q} that interpolates them to obtain the final answer as $\langle l_1,l_2\rangle$.

\medskip
\noindent\textbf{{\emph{User tracing} (lines~\ref{ln:function_user_trace_ss}-\ref{ln:user_notification_ss}).}} First, \textbf{Q} executes $\mathit{Location\_Trace}()$ to know the impacted locations by the infected person (line~\ref{ln:user_call_location_trace_ss}). Then, \textbf{Q} creates SSS of all impacted locations (denoted by $\mathit{sssLoc}[]$) and sends them to the servers along with the desired epoch-identifer (which is the same as used when knowing infected locations in line~\ref{ln:user_call_location_trace_ss}). \textbf{S} executes string-matching operation over each value of $A_{\mathit{smL}}$ against each value of $\mathit{sssLoc}[]$ in the desired epoch, and it will result in either 0 or 1 of secret-shared form). Then, the $i^{\mathit{th}}$ result of string-matching operation is multiplied by $i^{\mathit{th}}$ value of $A_{\mathit{sid}}$, resulting in the secret-shared device-ids, if impacted by the infected person; otherwise, the secret-shared location value will become 0 of secret-shared form (line~\ref{ln:s_find_impacted_person_ss}). Finally, \textbf{Q} receives shares of all impacted people from all servers, interpolates them, and it results in all impacted people (line~\ref{ln:user_get_impacted_person_ss}). All such impacted users are notified using $\mathit{Notification(\ast)}$ function of Algorithm~\ref{alg:Query Execution Algorithm}.

\noindent
\textit{\underline{Example~\ref{sec:Information-Theoretically Secure Solution}.2.}} We continue from Example~\ref{sec:Information-Theoretically Secure Solution}.1, where $d_1$ was the device of an infected person in Table~\ref{tab:Secret-shared WiFi relation for an epoch} and impacted locations were $\langle l_1,l_2\rangle$ that were known to \textbf{Q} after executing $\mathit{Location\_Trace}(\ast)$ (line~\ref{ln:function_location_trace_ss}). Now, to find impacted people, \textbf{Q} generates SSS of $l_1$ and $l_2$, say $\gamma_1$ and $\gamma_2$, respectively. \textbf{S} checks $\gamma_1$ and $\gamma_2$ against the four shares (via string-matching operation) in $A_{\mathit{smL}}$. It will result in two vectors: $\langle1,0,1,1\rangle$ of secret-shared form corresponding to $\gamma_1$ and $\langle0,1,0,0\rangle$ of secret-shared form corresponding to $\gamma_2$. Then, the vectors are position-wise multiplied by  $\langle\mathrm{S}(d_1),\mathrm{S}(d_2),\mathrm{S}(d_1),\mathrm{S}(d_1)\rangle$. Thus, \textbf{S} sends $\langle d_1,0,d_1,d_1\rangle$ and $\langle 0,d_2,0,0\rangle$ of secret-shared form to \textbf{Q}. \textbf{Q} interpolates the vectors and knows that the device $d_2$ belongs to an impacted person.

\medskip
\noindent\textbf{{\emph{Social distancing} (lines~\ref{ln:function_social_distance_ss}-\ref{ln:social_distance_issue_alarm_ss}).}} \textbf{Q} sends the desired epoch identifier to the servers (line~\ref{ln:quest_sends_delta}).
Based on the desired identifier, each server multiplies the $i^{\mathit{th}}$ value of $A_{\mathit{su}}$ with the $i^{\mathit{th}}$ value of $A_{\mathit{sL}}$, and it results in all locations having the unique devices. The server sends all such locations to \textbf{Q} (line~\ref{ln:social_server}).
First, \textbf{Q} interpolates the received locations (line~\ref{ln:q_interpolate_social_dis_location_ss}) and then, counts the appearance of each location (line~\ref{ln:q_groupby_social_dis_ss}). Finally, \textbf{Q} issues an alarm, if the counter value for a location exceeds the predefined rule for social distancing, denoted by $\mathrm{distanceIndex}$ (line~\ref{ln:social_distance_issue_alarm_ss}).

\textit{Aside.} Note that we can also know the devices that do not follow the predefined rule for social distancing, by multiplying the $i^{\mathit{th}}$ value of $A_{\mathit{su}}$ with the $i^{\mathit{th}}$ value of $A_{\mathit{sid}}$ in the desired epoch, and in experiment section \S\ref{sec:Experimental Evaluation}, we will find all such device-ids in our experiments.

\noindent
\textit{\underline{Example~\ref{sec:Information-Theoretically Secure Solution}.3.}} Suppose that $\mathrm{distanceIndex}=1$, \textit{i}.\textit{e}., if there are more than one person at a location during a given epoch, then it shows that people at the location are not following the predefined social distancing rules. Suppose all four tuples of Table~\ref{tab:Secret-shared WiFi relation for an epoch} belongs to an epoch. \textbf{S} executes  position-wise multiplication and send the output of the following to \textbf{Q}: $\langle S(1)\times S(l_1), S(1)\times S(l_2), S(1)\times S(l_2),S(0)\times S(l_1)\rangle$. \textbf{Q} interpolates the received answers, counts the number of each location (as $l_1=1$ and $l_2=2$), and finds that the location $l_2$ is not following the social distancing rule.

\medskip
\noindent
\textbf{Information leakage discussion.}  Since Algorithm~\ref{alg:Secret-share creation algorithm} uses different polynomials of the same degree for creating shares of a secret, an adversary cannot learn anything by observing the shares. Algorithm~\ref{alg:iquest_Query Execution Algorithm_sss} creates secret-shares of a query predicate that appears different from the secret-shared data. Thus, the adversary by observing the query predicate cannot learn which tuples satisfy the query. Furthermore, since Algorithm~\ref{alg:iquest_Query Execution Algorithm_sss} performs an identical operation on each share (\textit{e}.\textit{g}., lines~\ref{ln:server_trace_location_ss},\ref{ln:s_find_impacted_person_ss},\ref{ln:social_server}), it hides access-patterns; thus, the adversary cannot learn anything from the query execution, also. Hence, in \textsc{iQuest} provides stronger security than \textsc{cQuest}.

\medskip
\noindent\textbf{Pros.}  Due to hiding access-patterns, \textsc{iQuest} provides stronger security and satisfies the security properties given in~\S\ref{subsec:Security Properties}, \textit{i}.\textit{e}., query and execution privacy. Also, it prevents a computationally efficient adversary to know anything from the ciphertext. Also, it is fault-tolerant, due to using multiple servers.


\medskip
\noindent\textbf{Cons.}
As known, a fully secure system incurs performance overheads. Due to executing an identical operation on each share, and hence, not using any index structure, \textsc{iQuest} incurs the computational cost at the server. Also, since the servers send a secret-shared vector (having 0 or the desired value) of size equals to the numbers of tuples in the desired epoch, it incurs the communication cost. Nevertheless, \S\ref{sec:Experimental Evaluation}) will show that such overheads are not very high.

\medskip
\noindent\textbf{Optimization.} In social distancing application using \textsc{iQuest}, it may turn out that we need to send a significant amount of data from the servers to \textsc{Quest}. To avoid such communication, \textbf{Q} can send SSS of all the locations (denoted by $\mathit{smLoc}[]$) to servers. The servers can do the following:

\centerline{
$\mathit{count}_{i} \leftarrow \Sigma_{1\leq j\leq y}(A_{\mathit{smL}}[j]\otimes \mathit{sssLoc}[i])\times A_{\mathit{su}}[j], \forall i\in \{1,|\mathit{smLoc}[]|\}$
}

The servers execute string-matching operation for each location of $\mathit{smLoc}[]$ against each value of $A_{\mathit{smL}}$ in each desired epoch and adds the output of string matching operation. Thus, for each desired epoch, the servers send $|\mathit{sssLoc}[]|$ numbers to \textbf{Q} that interpolates them to know the number of unique devices at each location in the desired epochs. Note that this method will outperform the method given in Algorithm~\ref{alg:iquest_Query Execution Algorithm_sss} Line~\ref{ln:join_operation}, if the number of tuples in each epoch are more than the number of locations in $\mathit{smLoc}[]$.

\begin{figure*}[!t]
		\begin{center}
			\begin{minipage}{.24\linewidth}
				\centering
	    \includegraphics[scale=0.28]{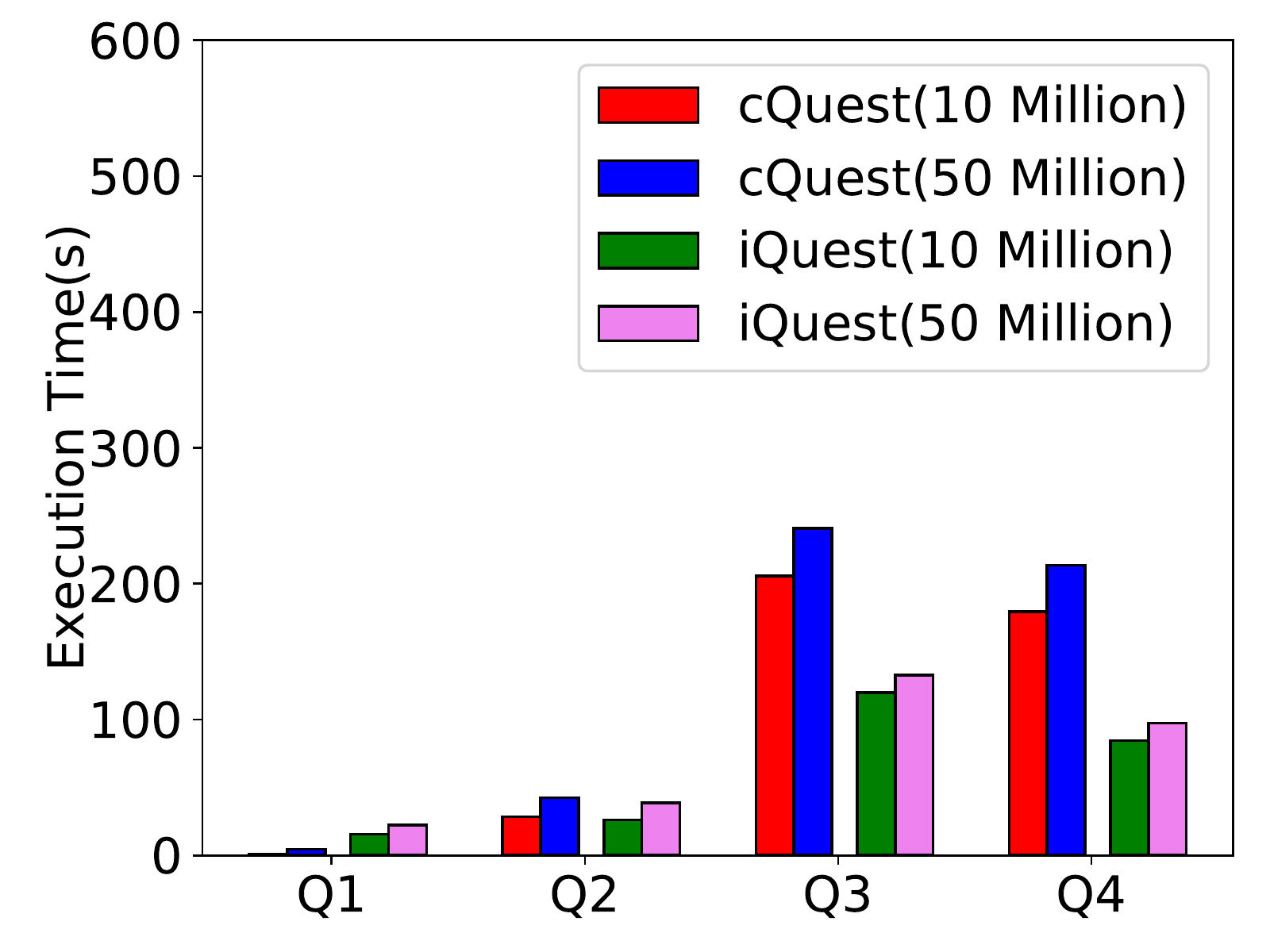}
        \subcaption{1-infected, 1-day.}
        \label{fig:3a}
			\end{minipage}
			\begin{minipage}{.24\linewidth}
				\centering
	    \includegraphics[scale=0.28]{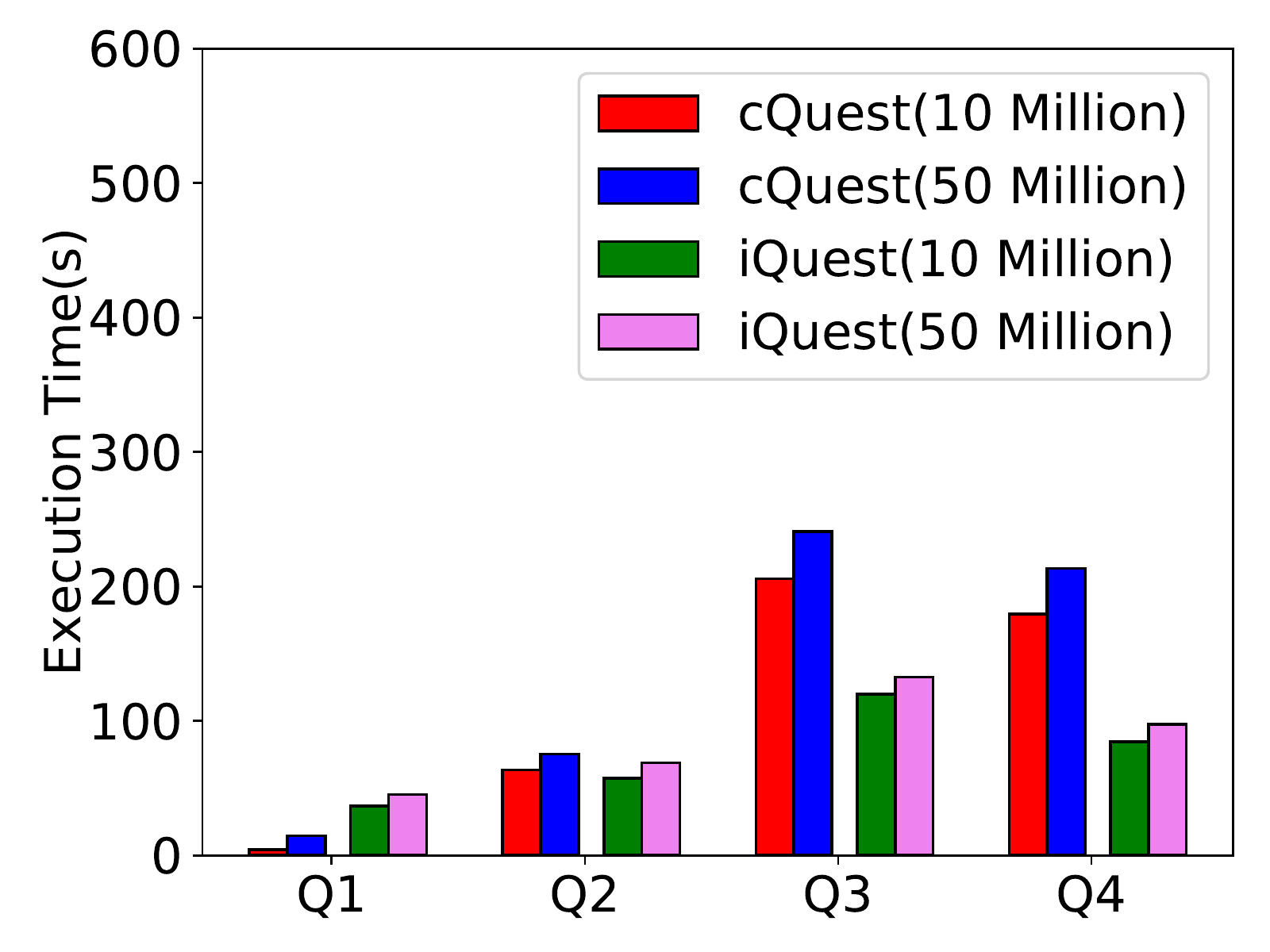}
               \subcaption{100-infected, 1-day.}
        \label{fig:3b}
			\end{minipage}			
		\begin{minipage}{.24\linewidth}
				\centering
            	\includegraphics[scale=0.28]{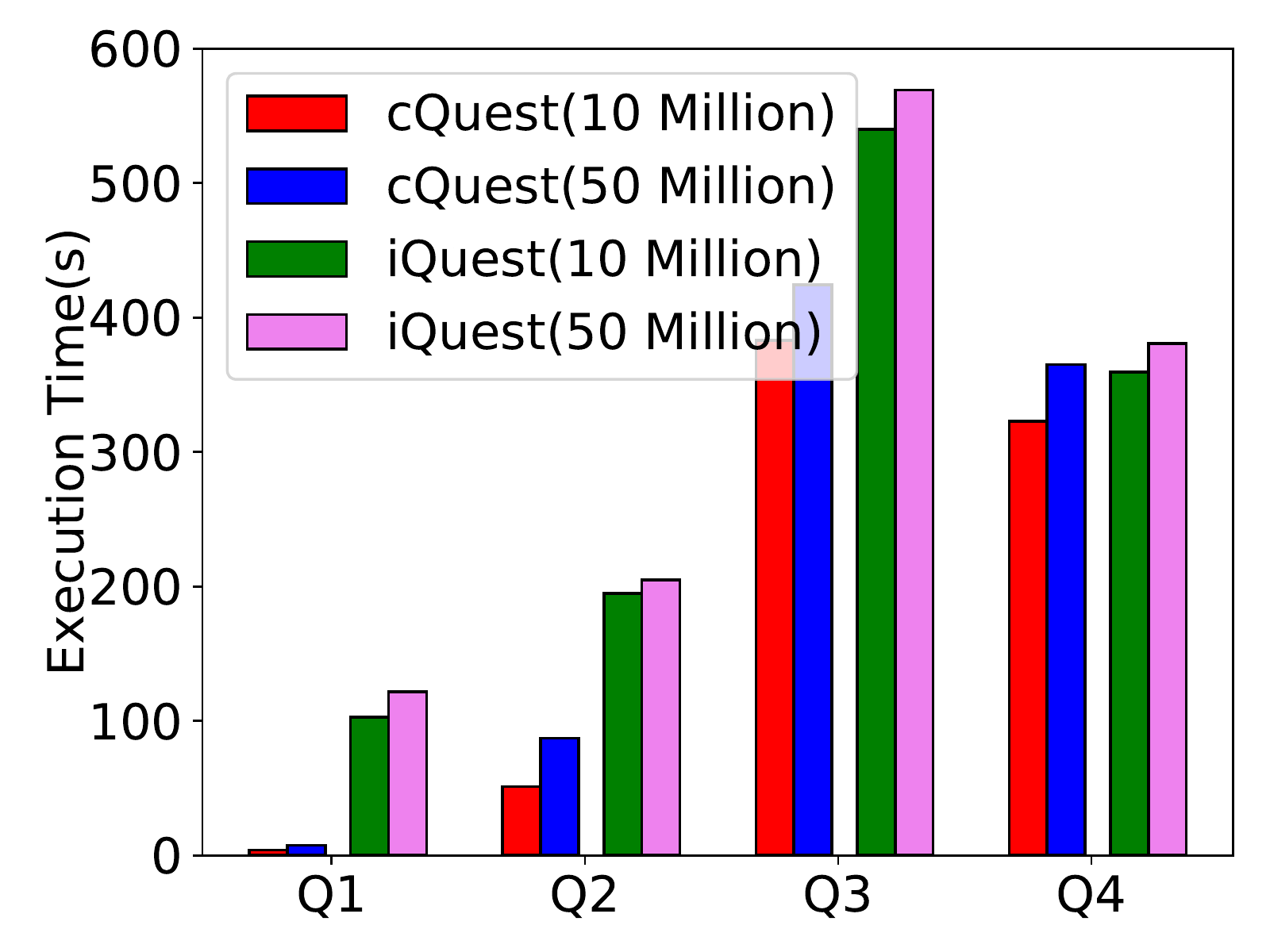}
                \subcaption{1-infected, 14-days.}
                \label{fig:3c}
			\end{minipage}
			\begin{minipage}{.24\linewidth}
				\centering
            	\includegraphics[scale=0.28]{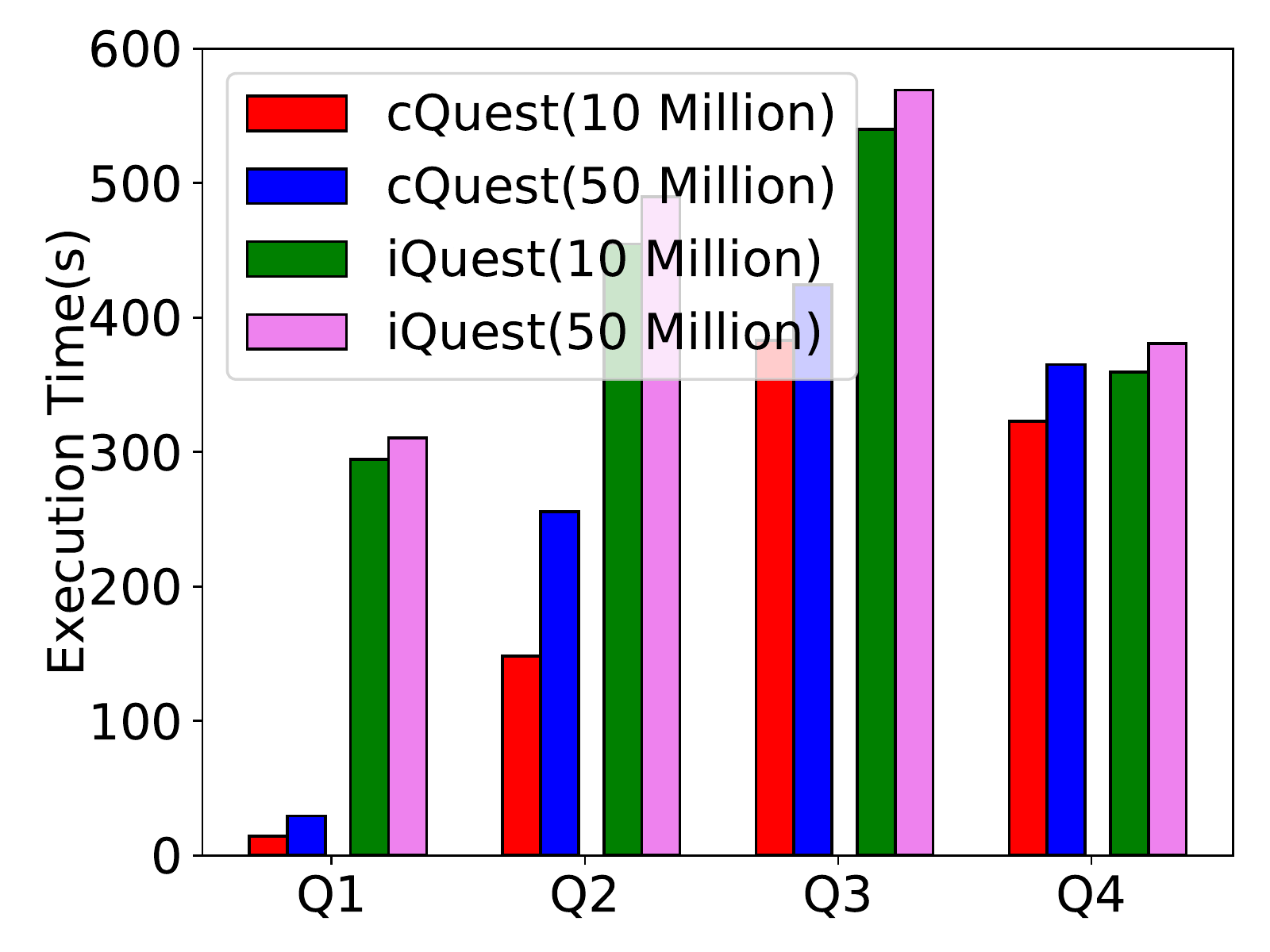}
                \subcaption{100-infected, 14-days.}
                \label{fig:3d}
            \end{minipage}
		\end{center}
		\caption{Exp 3: Scalability test of 10M and 50rows with varying other parameters.}
		\label{fig:Scalability test}
	\end{figure*}

\begin{figure*}[!t]
		\begin{center}
			\begin{minipage}{.24\linewidth}
				\centering
	    \includegraphics[scale=0.28]{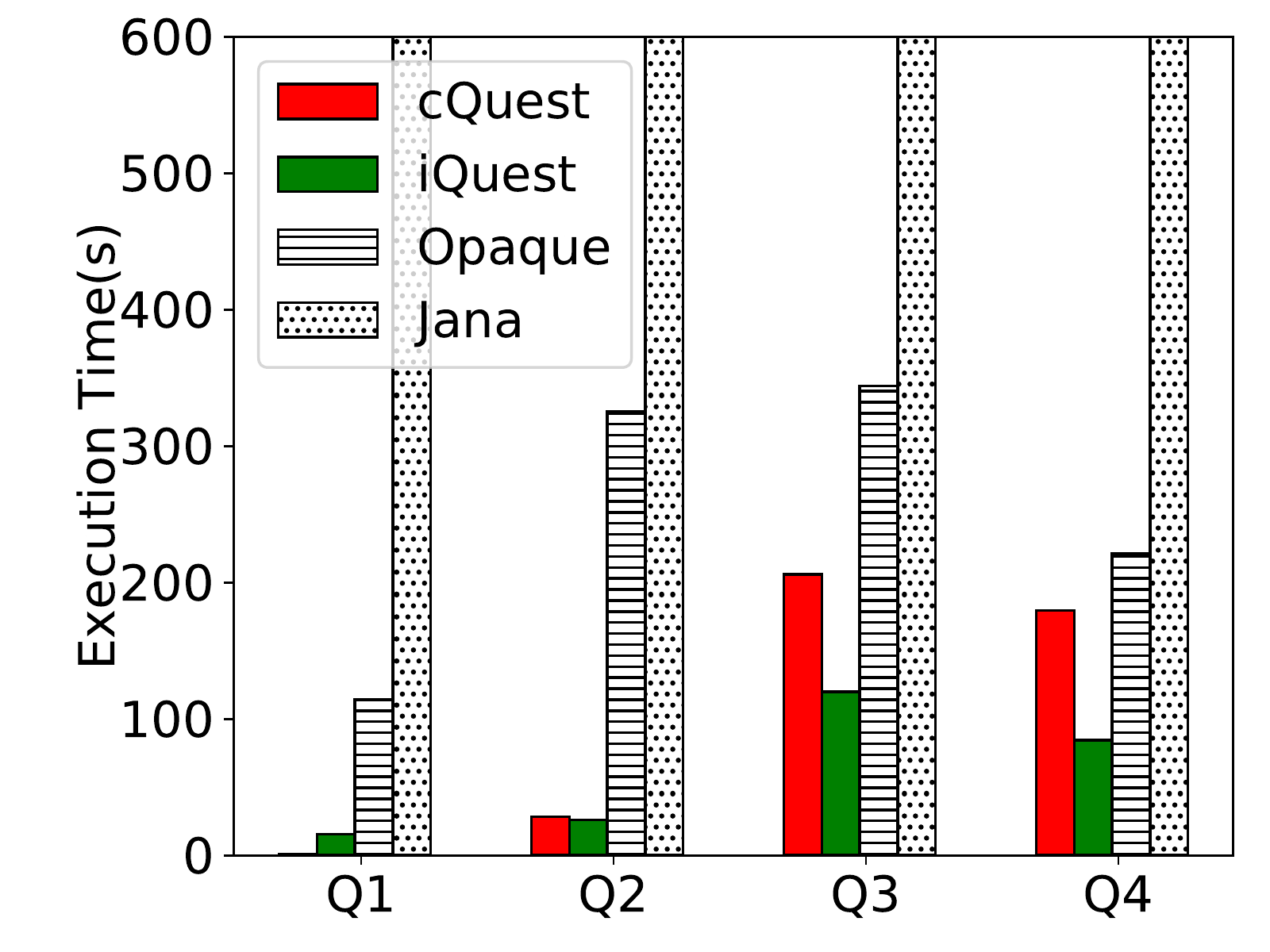}
        \subcaption{1-infected, 1-day.}
        \label{fig:4a}
			\end{minipage}
			\begin{minipage}{.24\linewidth}
				\centering
	    \includegraphics[scale=0.28]{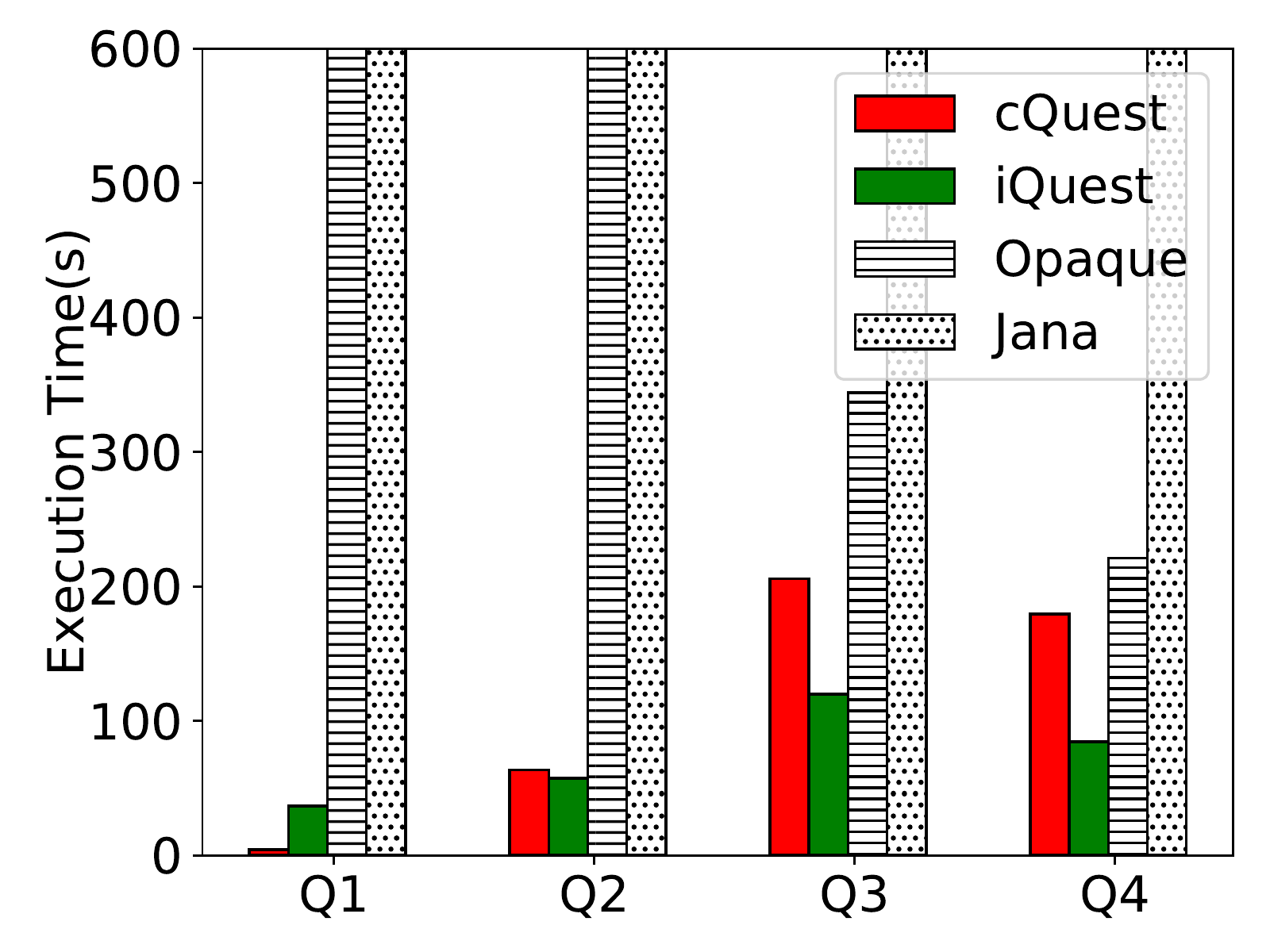}
        \subcaption{100-infected, 1-day.}
        \label{fig:4b}
			\end{minipage}			
		\begin{minipage}{.24\linewidth}
				\centering
            	\includegraphics[scale=0.28]{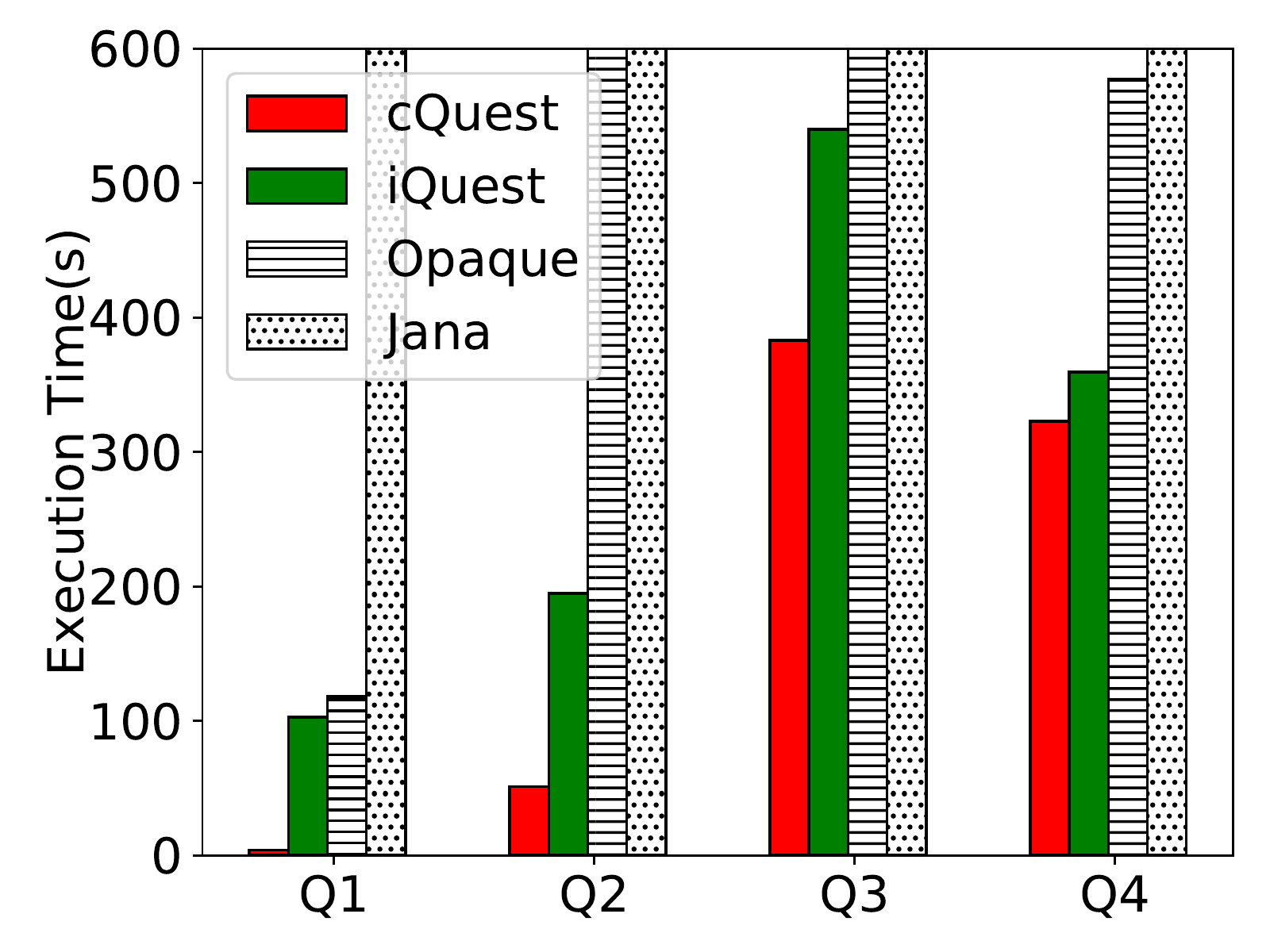}
                \subcaption{1-infected, 14-days.}
                \label{fig:4c}
			\end{minipage}
			\begin{minipage}{.24\linewidth}
				\centering
            	\includegraphics[scale=0.28]{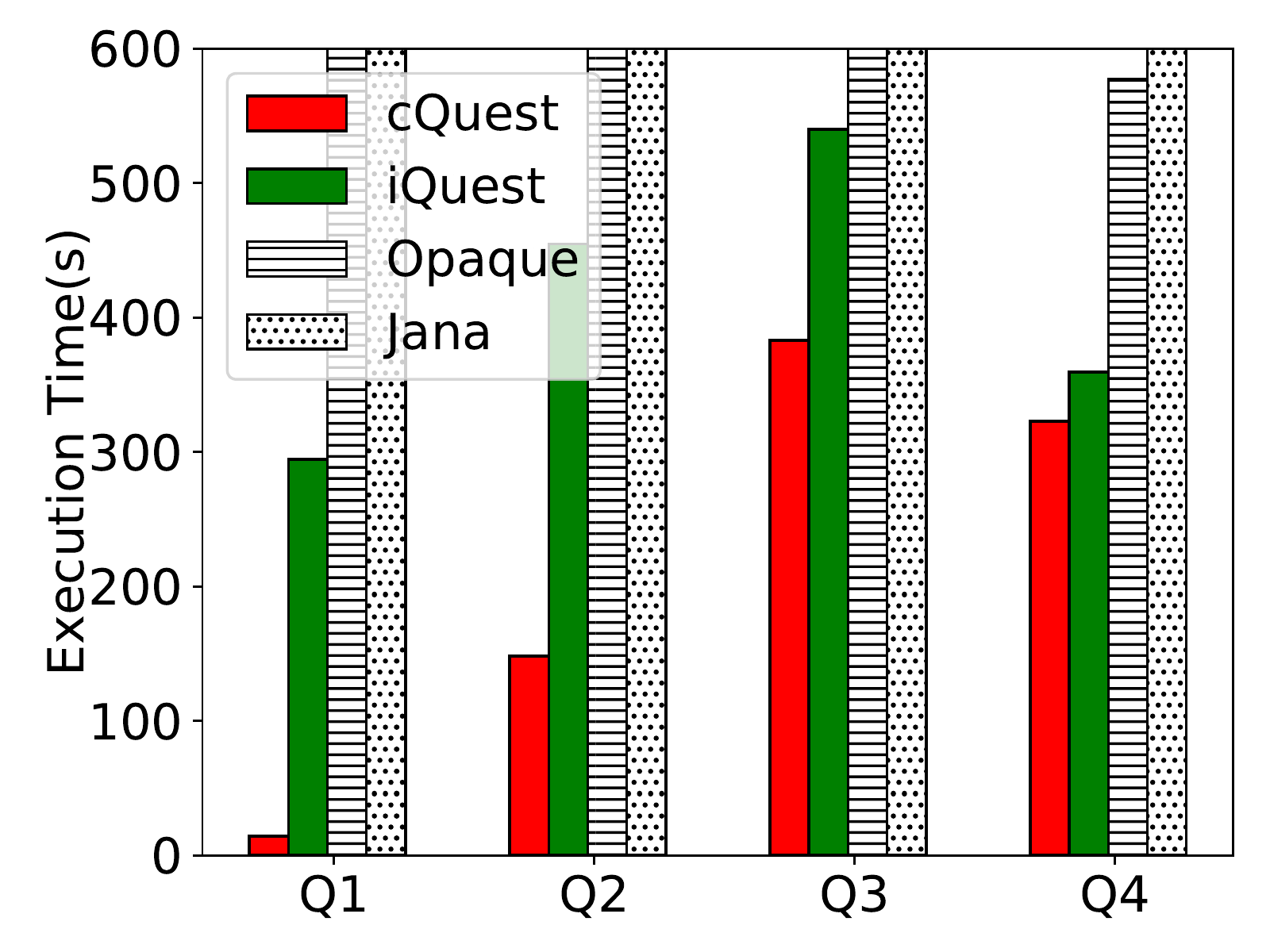}
                \subcaption{100-infected, 14-days.}
                \label{fig:4d}
            \end{minipage}
		\end{center}
\caption{Exp 4: Using other systems (secure hardware based Opaque and MPC-based Jana) vs \textsc{cQuest} and \textsc{iQuest} on 10M.}
		\label{fig:using other systems}
	\end{figure*}

\section{Experimental Evaluation}
\label{sec:Experimental Evaluation}
\textsc{Quest} is deployed at UCI, where it is being used to support social distancing and crowd flow applications~\cite{uciapp}. This section evaluates the scalability of \textsc{Quest} to evaluate its practicality for larger deployments and for all supported applications. We used AWS servers with 192GB RAM, 3.5GHz Intel Xeon CPU with 96 cores and installed MySQL to store secured dataset. A 16GB RAM machine at the local-side hosts \textsc{Quest} that communicates with AWS servers.

\smallskip
\noindent
\textbf{Dataset.} We used WiFi association data generated using SNMP traps at the campus-level WiFi infrastructure at UCI that consists of 2000 access-points with four controllers. Experiments used real-time data received at one of the four controllers (that collects real-time WiFi data from 490 access-points spread over 40+ buildings). Using this WiFi data, we created two types of datasets, refer to Table~\ref{tab:dataset properties}.
For evaluating \textsc{iQquest}, we created nine shares, since at most $2(x+y)+1$ shares are required, where $x=3$ (the length of device-ids, line~\ref{ln:find_last_v_bit_sss} Algorithm~\ref{alg:Secret-share creation algorithm}) and $y=1$ (a single secret value in column $A_{\mathit{sL}}$, line~\ref{ln:create_share_location_ss} Algorithm~\ref{alg:Secret-share creation algorithm}).


\bgroup
\def\arraystretch{.86}
\begin{table}[!h]
\centering
\scriptsize
\begin{tabular}{|l|l|l|l|l|l|l|l|}\hline
\#rows & Cleartext size & Days covered & Encrypted size  & Secret-Share size  \\\hline

  10M & 1.4GB & 14 & 5GB  & 25GB\\\hline
  50M & 7.0GB & 65 & 13GB & 65GB \\\hline

\end{tabular}
\caption{Characteristics of the datasets used in experiments.}
\label{tab:dataset properties}
\end{table}
\egroup

\smallskip
\noindent\textbf{Queries.} We executed all queries (Q1: social distancing, Q2: Contact tracing, Q4: Crowd-flow), see Table~\ref{tab:queries}. We modified `Q3: social distancing query' by also fetching device-ids that do not follow distancing rules, in addition to fetching locations information.

\medskip
\noindent
\textbf{Exp 1: Throughput.} In order to evaluate the overhead of \textsc{cQuest} and \textsc{iQuest} at the ingestion time, we measured the throughput (rows/minute) that \textsc{Quest} can sustain. \textsc{cQuest} Algorithm~\ref{alg:Data Encryption Algorithm} can encrypt $\approx$494,226 tuples/min, and \textsc{iQuest} Algorithm~\ref{alg:Secret-share creation algorithm} can create secret-shares of  $\approx$38,935 tuples/min. While the throughput of Algorithm~\ref{alg:Secret-share creation algorithm}'s is significantly less than Algorithm~\ref{alg:Data Encryption Algorithm}, as it needs to create 9 (different) shares, it can sustain UCI level workload on the relatively weaker machine used for hosting \textsc{Quest}.   

\medskip
\noindent
\textbf{Exp 2: Metadata size.} Recall that
Algorithm~\ref{alg:iquest_Query Execution Algorithm_sss} (Algorithm~\ref{alg:Query Execution Algorithm}) for \textsc{iQuest} (\textsc{cQuest}) maintains hash-tables for a certain duration. Table~\ref{tab:hash table size} shows the size of hash tables created for epochs of different sizes: 15min, 30min, and 60min. Note that the metadata size for \textsc{cQuest} is larger than \textsc{iQuest}, since \textsc{cQuest} uses two hash tables (line~\ref{ln:new_hash_table_location} Algorithm~\ref{alg:Query Execution Algorithm}) and one list of visited places by each device, while \textsc{iQuest} uses only one hash table (line~\ref{ln:new_hash_table_device_device_sss} Algorithm~\ref{alg:iquest_Query Execution Algorithm_sss}) and the list.
Metadata overheads remain small for both techniques.

\bgroup
\def\arraystretch{1}
\begin{table}[!h]
\centering
\begin{tabular}{|l|l|l|l|l|}\hline

Epoch duration & \textsc{cQuest} & \textsc{iQuest} \\\hline
  15min & 1.96MB & 0.93MB \\\hline
  30min & 3.40MB & 1.37MB \\\hline
  60min & 5.84MB & 2.10MB \\\hline

\end{tabular}
\caption{Exp 2: Size of hash tables, for different epoch sizes.}
\label{tab:hash table size}
\end{table}
\egroup

\medskip
\noindent
\textbf{Exp 3: Scalability.} We measured the scalability of \textsc{Quest} in three scenarios, by varying the number of infected people, days for tracing, and dataset size. Figure~\ref{fig:Scalability test} shows results for 1-100 infected users for Q1, Q2 and execution of Q1-Q4 over 1-14 days duration on 10M, 50M rows. In Q1, a device has visited between 1 to 55 locations in 1 epoch. Note that Q1 using \textsc{cQuest} took less time in all four cases, since it uses an index on $A_{\mathit{id}}$ column (line~\ref{ln:fetch_row_using_index_location_encryption} Algorithm~\ref{alg:Query Execution Algorithm}); while \textsc{iQuest} took more time, since it scans all data depending on the queried  interval (line~\ref{ln:server_trace_location_ss} Algorithm~\ref{alg:iquest_Query Execution Algorithm_sss}). As the number of infected people increases, the query time increases too. Cost analysis follows the same argument as Q2 that is an extension of Q1.


For Q3 and Q4 in Figures~\ref{fig:3a} and~\ref{fig:3b}, \textsc{iQuest} took less time than \textsc{cQuest}. The reason is: \textsc{iQuest} performs multiplication on $i^{\mathit{th}}$ values of $A_{\mathit{sL}}$ and $A_{\mathit{\mathit{su}}}$ (line~\ref{ln:social_server} Algorithm~\ref{alg:iquest_Query Execution Algorithm_sss}), and the cost depends on the number of tuples in the desired epochs. However, \textsc{cQuest} joins a table of size $y\times\Delta_t\times x$ with the encrypted WiFi data table on $A_{L}$ column to obtain the number of locations having unique devices (line~\ref{ln:join_operation} Algorithm~\ref{alg:Query Execution Algorithm}), where $y$ is the maximum appearance of a location in any epoch (can be of the order of 10,000, causing a larger join table size), $\Delta_t$ is the number of desired epochs, and $x$ is the number of locations. Also, note that for Q3 and Q4 in Figures~\ref{fig:3c} and~\ref{fig:3d}, \textsc{iQuest} took more time than \textsc{cQuest}, since the increase in the cost of multiplication operations (due to larger dataset of 14-days tracing period) in \textsc{iQuest} overtook the increase in the cost of join in \textsc{cQuest}. It shows  \textsc{cQuest} is more scalable than \textsc{iQuest}.

\medskip
\noindent
\textbf{Exp 4: Using other existing systems to support \textsc{Quest} applications.}
We note an \textbf{alternative solution}, where one may output non-deterministically encrypted~\cite{DBLP:journals/jcss/GoldwasserM84} or secret-shared WiFi data (via \textsc{Quest}'s encrypter), on which the queries can be executed using existing SSEs~\cite{DBLP:journals/pvldb/LiLWB14,DBLP:conf/icde/LiL17}, secure hardware-based systems, \textit{e}.\textit{g}., Opaque~\cite{opaque}, or MPC-systems. Note that this solution does not need to develop any encryption or query execution algorithm. However, it may allow an adversary to deduce the user locations by observing datasets belonging to different organizations and \emph{may incur the high computational cost}, as will be clear below. 

Thus, to see the impact of using existing systems to support \textsc{Quest} applications, we used SGX-based Opaque~\cite{opaque} and MPC-based Jana~\cite{jana} on 10M rows only (since these systems were available to us and work on any dataset). Now, we can compare \textsc{cQuest} against computationally-secure Opaque and \textsc{iQuest} against information-theoretically-secure Jana. We inserted data, using non-deterministic encryption in Opaque and using the underlying secret-sharing mechanism in Jana. Then, we used their query execution mechanisms for queries Q1-Q4. Figure~\ref{fig:using other systems} shows the impact of using different systems for supporting our four queries on 10M rows. We drop any query that took more than 1000s.  

Observe that \textsc{cQuest} works well compared to Opaque, since \textsc{cQuest} uses index-based retrieval, while Opaque reads entire data in secure memory and decrypts it. \textsc{cQuest} and Opaque provides \emph{the same security}, \textit{i}.\textit{e}., ciphertext indistinguishability, and reveals access-patterns. Note that \textsc{cQuest} reveals access-patterns via index-scan, while Opaque reveals access-patterns due to side-channel (cache-line~\cite{DBLP:conf/eurosec/GotzfriedESM17} and branch-shadow~\cite{DBLP:conf/ccs/WangCPZWBTG17}) attacks.
Also, \textsc{iQuest} is efficient compared to Jana that takes more than 1000s in each query. The reason is: \textsc{iQuest} does not require communication among servers due to using string-matching over secret-shares~\cite{DBLP:conf/ccs/DolevGL15}, while Jana requires communication among servers, since Jana is based on MPC techniques that require communication among server during a computation to compute the answer. But, \textsc{iQuest} and Jana provide \emph{identical security} by hiding access-patterns, due to executing identical operations on each tuple.

\medskip
\noindent
\textbf{Exp 5: Impact of optimization.} We implemented improved methods to minimize the value of max location counter (\S\ref{sec:Computationally-Secure Solution}) and measured the performance improvement over 10M rows, while fixing the number of infected people to 100 and interval duration to 1-day. When we used \emph{counter per epoch} for Q2, it reduced the computation time from 63 (Figure~\ref{fig:3b}) to $\approx$35s and used 128KB more space to maintain the counter; while using \emph{counter per epoch and per location}, Q2 took only $\approx$2sec with 55MB space to store the counters.

We also implemented the improved method for uniqueness finding by outsourcing encrypted hash tables for each epoch. It reduced the time of Q4 (that finds unique devices in each epoch) from 179.4s to 1s. Further, we incorporated this improved method in Q3 (that also finds the devices that does not follow social distancing rule) with \emph{counter per epoch and per location} optimization, and it reduced the time of Q3 from 206s to 2s.

\begin{figure}[!t]
		\begin{center}
			\begin{minipage}{.49\linewidth}
				\centering
	    \includegraphics[scale=0.28]{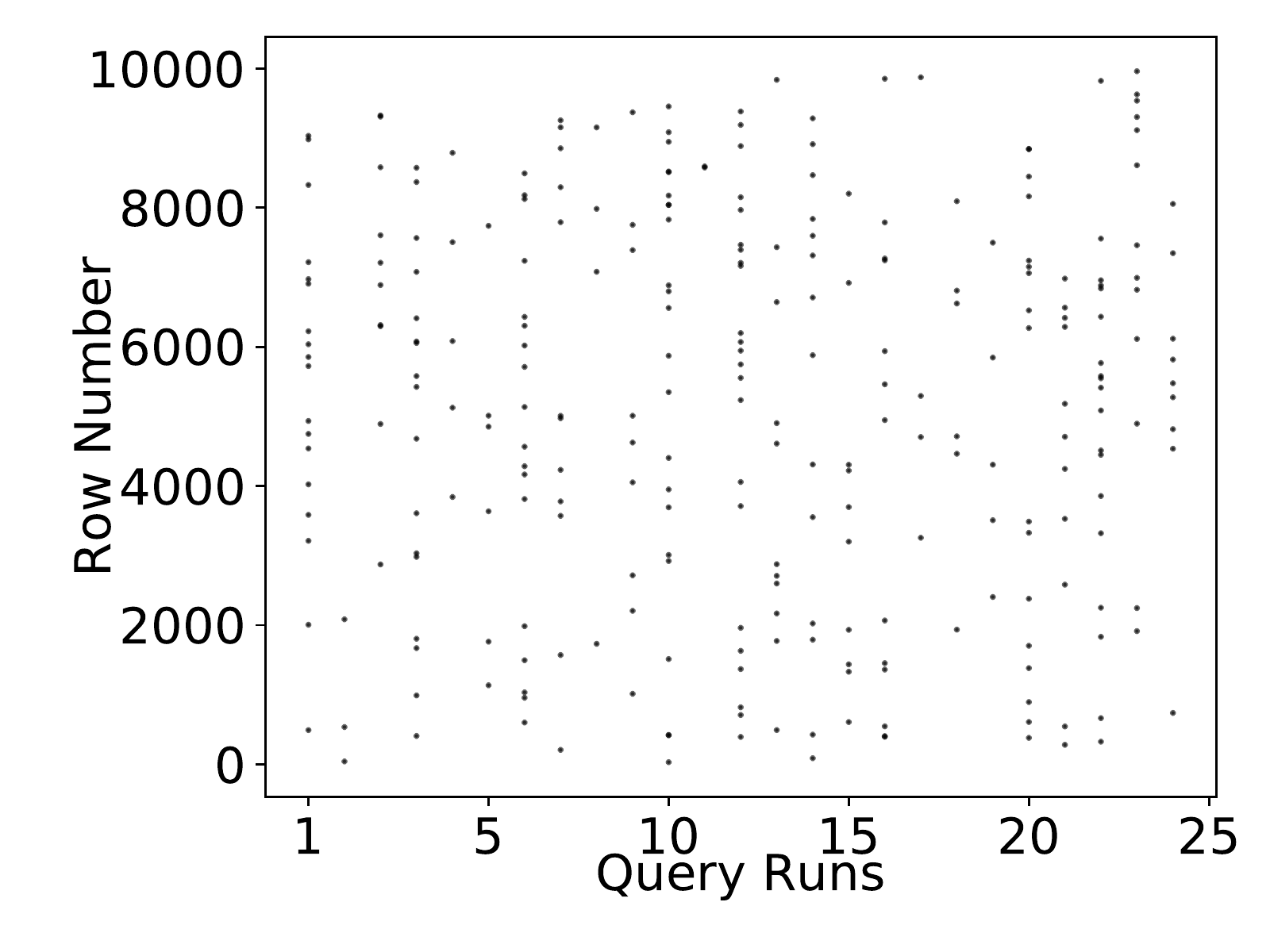}
        \subcaption{Access-patterns of \textsc{cQuest}.}
        \label{fig:fig_ap_cquest}
			\end{minipage}
			\begin{minipage}{.49\linewidth}
				\centering
	    \includegraphics[scale=0.28]{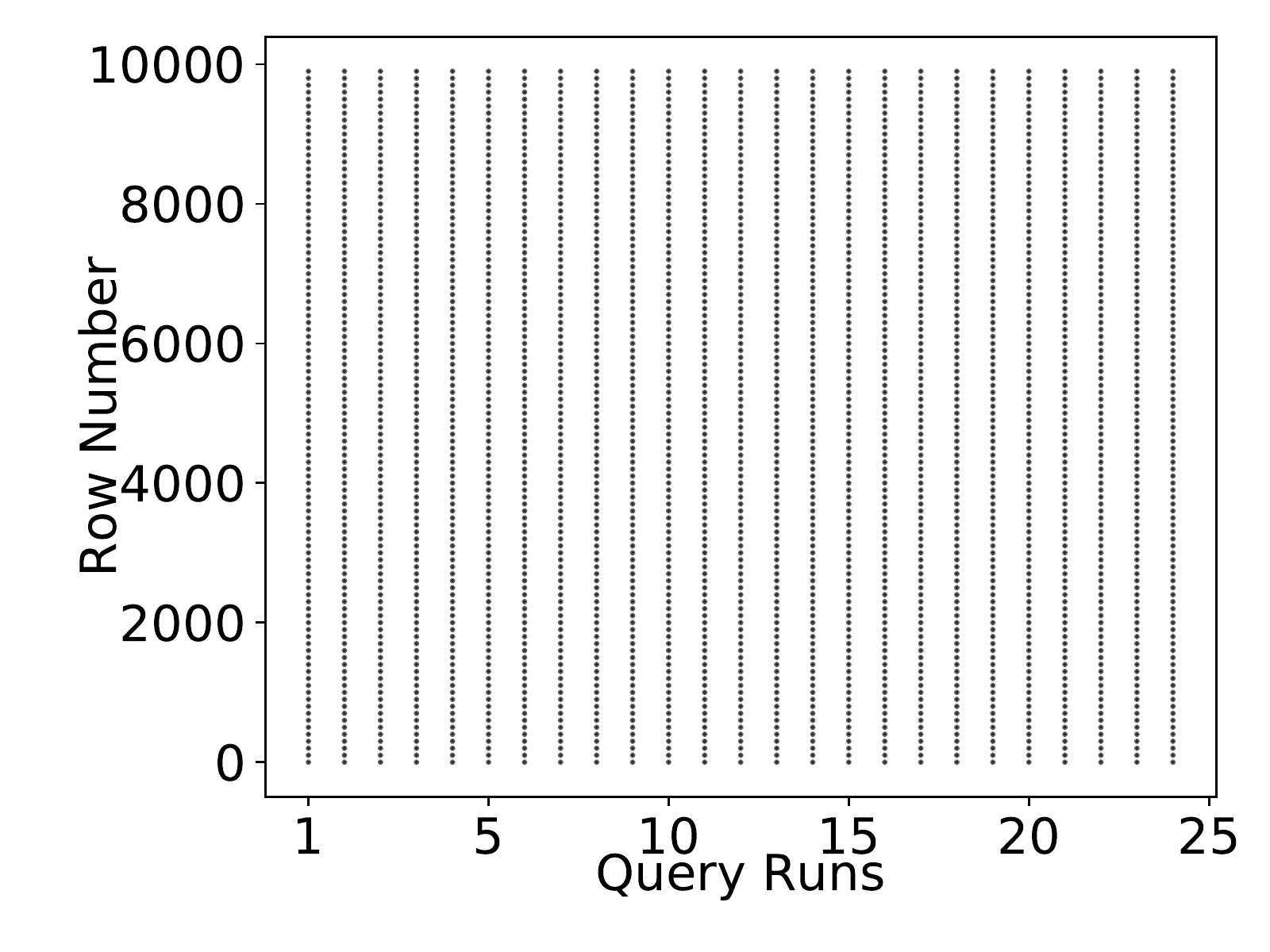}
        \subcaption{Access-patterns of \textsc{iQuest}.}
        \label{fig:fig_ap_iquest}
			\end{minipage}		
\caption{Exp 6: Access-patterns created by \textsc{Quest}.}
		\label{fig:fig_ap_quest}
\end{center}
	\end{figure}

\medskip
\noindent
\textbf{Exp 6: Access-patterns.} Figure~\ref{fig:fig_ap_quest} shows a sequence of memory accesses by \textsc{cQuest} and \textsc{iQuest}. For this, we run Q2 multiple times, selecting different device-ids each time over a fixed set of epochs. It is clear that
\textsc{iQuest} accesses the same memory locations (accesses all the rows of the given set of epochs) and produces an output for each accessed row for different queries, while \textsc{cQuest} accesses different memory locations (different rows for different device-ids) for answering different queries.

\medskip
\noindent
\textbf{Exp 7: Impact of communication.} Table~\ref{tab:cquest amount of data transfer} shows the amount of data transfer using \textsc{cQuest} and the data transfer time using different transfer speeds. From Table~\ref{tab:cquest amount of data transfer}, it is clear that \textsc{cQuest} is communication efficient, while \textsc{cQuest} reveals information from access-patterns. In particular, without using optimization methods (as described in \S\ref{sec:Computationally-Secure Solution}), Q3 and Q4 incur significant communication overheads, \textit{i}.\textit{e}., fetch $\approx$95MB data from the server. However, the optimization methods reduce such data size to $\approx$57KB.

\bgroup
\def\arraystretch{1}
\begin{table}[!h]
\centering
\begin{tabular}{|l|l|l|l|l|}\hline

Criteria             & Q1   & Q2             & Q3     & Q4   \\\hline
Without optimization & 1.4KB& 42.2KB         & 95MB   & 95MB  \\\hline
With optimization    & N/A  & N/A            & 56.6KB & 14.4KB \\\hline
Trans. speed 25MB/s  & Neg. & Neg.  & $\approx$2.5m    & Neg. \\\hline
Trans. speed 100MB/s & Neg. & Neg.   & $\approx$1m    & Neg. \\\hline
Trans. speed 500MB/s & Neg. & Neg.   & $\approx$11s    & Neg. \\\hline

\end{tabular}
\caption{Exp 7: \textsc{cQuest}: amount of data transfer and required time (Neg. refers to negligible).}
\label{tab:cquest amount of data transfer}
\end{table}
\egroup

Table~\ref{tab:iquest amount of data transfer} shows the amount of data transfer using \textsc{iQuest} and the data transfer time using different transfer speeds. From Table~\ref{tab:iquest amount of data transfer}, it is clear that \textsc{iQuest} incurs communication overhead, while \textsc{iQuest} provides a high-level of security. In particular, Q1 requires us to fetch $\approx$32MB data from each server when tracing period was 14-days for an infected person. As Q2 requires two communication rounds (the first for knowing the impacted location and another for knowing the impacted device ids), it incurs signification communication cost by fetching $\approx$3.5GB data from each server. The reason is: we need to fetch data corresponding to 55 locations that a user can visit during an epoch. Q3, also, incurs the same communication overhead. Q4 requires to download $\approx$32MB data from each server for executing social distancing over 14-days. However, when we use the improved method (as described in \S\ref{sec:Information-Theoretically Secure Solution}) for Q4, we need to fetch only 2.1MB data. 

\bgroup
\def\arraystretch{1}
\begin{table}[!h]
\centering
\begin{tabular}{|l|l|l|l|l|}\hline

Criteria             & Q1   & Q2             & Q3     & Q4   \\\hline
Without optimization & 32MB & 3.6GB          & 3.6GB  & 32MB  \\\hline
With optimization    & N/A  & N/A            & N/A    & 2.1MB \\\hline
Trans. speed 25MB/s  & Neg. & $\approx$2.5m  & $\approx$2.5m    & Neg. \\\hline
Trans. speed 100MB/s & Neg. & $\approx$1m    & $\approx$1m    & Neg. \\\hline
Trans. speed 500MB/s & Neg. & $\approx$11s   & $\approx$11s    & Neg. \\\hline

\end{tabular}
\caption{Exp 7: \textsc{iQuest}: amount of data transfer and required time (Neg. refers to negligible).}
\label{tab:iquest amount of data transfer}
\end{table}
\egroup

%

\section{Lessons Learnt}

In this paper, we designed, developed, and validated a system, called \textsc{Quest} for privacy-preserving presence and contact tracing at the organizational level using WiFi connectivity data  to enable community safety in a pandemic. \textsc{Quest} incorporates a flexible set of methods that can be customized depending on the desired privacy needs of the smartspace and its associated data. We anticipate that capabilities provided by  \textsc{Quest} are vital for organizations to resume operations after a community-scale lockdown --- the passive approach to information gathering in \textsc{Quest} can enable  continuous information awareness to encourage social distancing measures and identify settings and scenarios, where additional caution should be exercised. Ongoing discussions with campus administration at UC Irvine to utilize \textsc{Quest}'s capabilities for a staged and guided reopening of campus have highlighted the value of the privacy and security features embedded in \textsc{Quest}. The living lab experience at UC Irvine will enable us to tune the underlying cryptographic protocols for other useful applications including dynamic occupancy counts and context-aware messaging to encourage safe operations.


\begin{thebibliography}{10}

\bibitem{google-apple}
Apple's and Google's COVID-19 contact tracing technology, available at:
  \url{https://tinyurl.com/wfw9ojr}.

\bibitem{pepppt}
Pan-European Privacy-Preserving Proximity Tracing: available at:
  \url{https://www.pepp-pt.org/}.

\bibitem{IsraelsTheShield}
Israel's The Shield: available at: \url{https://tinyurl.com/y75bqjj9}.

\bibitem{TraceTogether}
TraceTogether, available at: \url{https://www.tracetogether.gov.sg/}.

\bibitem{SouthKorea100m}
South Korea's 100m: available at: \url{https://tinyurl.com/yb5mj9o6}.

\bibitem{cnn}
Georgia's daily coronavirus deaths will nearly double by August with relaxed
  social distancing, model suggests, available at:
  \url{https://tinyurl.com/ydy53cfc}.

\bibitem{vox}
Polls: Americans don’t want to end social distancing policies despite
  financial devastation, available at: \url{https://tinyurl.com/ybvtfn9a}.

\bibitem{uciapp}
\textsc{Quest} Applications: available at:
  \url{https://tippersweb.ics.uci.edu/covid19/d/IwAc1O9Wk/covid-19-effort-at-uc-irvine?orgId=1}.

\bibitem{stanford}
Stanford University's COVID-Watch, available at:
  \url{https://covid-watch.org/}.

\bibitem{fakenews1}
Fake news about the coronavirus is hazardous to your health. Don't fall for it:
  Doctor, available at: \url{https://tinyurl.com/ybk4b5lo}.

\bibitem{fakenews2}
During this coronavirus pandemic, `fake news' is putting lives at risk: UNESCO,
  available at: \url{https://tinyurl.com/y78jhdbl}.

\bibitem{enigma}
SafeTrace, available at: \url{https://github.com/enigmampc/safetrace}.

\bibitem{DBLP:conf/sigmod/AgrawalKSX04}
R.~Agrawal et~al.
\newblock Order-preserving encryption for numeric data.
\newblock In {\em {SIGMOD}}, pages 563--574, 2004.

\bibitem{DBLP:journals/corr/abs-2004-04145}
A.~Aktay et~al.
\newblock Google {COVID-19} community mobility reports: Anonymization process
  description (version 1.0).
\newblock {\em CoRR}, abs/2004.04145, 2020.

\bibitem{DBLP:conf/icc/AltuwaiyanHL18}
T.~Altuwaiyan et~al.
\newblock {EPIC:} efficient privacy-preserving contact tracing for infection
  detection.
\newblock In {\em {ICC}}, pages 1--6, 2018.

\bibitem{DBLP:conf/eurosec/AmjadKM19}
G.~Amjad et~al.
\newblock Forward and backward private searchable encryption with {SGX}.
\newblock In {\em Proceedings of the 12th European Workshop on Systems
  Security, EuroSec@EuroSys 2019, Dresden, Germany, March 25, 2019}, pages
  4:1--4:6, 2019.

\bibitem{jana}
D.~W. Archer et~al.
\newblock From keys to databases - real-world applications of secure
  multi-party computation.
\newblock {\em Comput. J.}, 61(12):1749--1771, 2018.

\bibitem{DBLP:conf/crypto/BellareBO07}
M.~Bellare et~al.
\newblock Deterministic and efficiently searchable encryption.
\newblock In {\em {CRYPTO}}, pages 535--552, 2007.

\bibitem{DBLP:conf/percom/BiXHHPM17}
C.~Bi et~al.
\newblock Familylog: {A} mobile system for monitoring family mealtime
  activities.
\newblock In {\em {PerCom}}, pages 21--30, 2017.

\bibitem{DBLP:conf/crypto/BoldyrevaFO08}
A.~Boldyreva et~al.
\newblock On notions of security for deterministic encryption, and efficient
  constructions without random oracles.
\newblock In {\em {CRYPTO}}, pages 335--359, 2008.

\bibitem{DBLP:conf/eurocrypt/BrandsC93}
S.~Brands et~al.
\newblock Distance-bounding protocols (extended abstract).
\newblock In {\em {EUROCRYPT}}, pages 344--359, 1993.

\bibitem{DBLP:conf/stoc/CanettiFGN96}
R.~Canetti et~al.
\newblock Adaptively secure multi-party computation.
\newblock In {\em {STOC}}, pages 639--648, 1996.

\bibitem{canetti2020anonymous}
R.~Canetti, A.~Trachtenberg, and M.~Varia.
\newblock Anonymous collocation discovery:taming the coronavirus while
  preserving privacy, 2020.

\bibitem{cho2020contact}
H.~Cho, D.~Ippolito, and Y.~W. Yu.
\newblock Contact tracing mobile apps for covid-19: Privacy considerations and
  related trade-offs, 2020.

\bibitem{corless2013graduate}
R.~M. Corless and N.~Fillion.
\newblock A graduate introduction to numerical methods.
\newblock {\em AMC}, 10:12, 2013.

\bibitem{sgx}
V.~Costan and S.~Devadas.
\newblock Intel {SGX} explained.
\newblock {\em {IACR} Cryptology ePrint Archive}, 2016:86, 2016.

\bibitem{DBLP:journals/jcs/CurtmolaGKO11}
R.~Curtmola et~al.
\newblock Searchable symmetric encryption: Improved definitions and efficient
  constructions.
\newblock {\em Journal of Computer Security}, 19(5):895--934, 2011.

\bibitem{DBLP:journals/corr/abs-2003-13073}
D.~Demirag et~al.
\newblock Tracking and controlling the spread of a virus in a
  privacy-preserving way.
\newblock {\em CoRR}, abs/2003.13073, 2020.

\bibitem{DBLP:conf/ccs/DolevGL15}
S.~Dolev et~al.
\newblock Accumulating automata and cascaded equations automata for
  communicationless information theoretically secure multi-party computation.
\newblock {\em Theor. Comput. Sci.}, 795:81--99, 2019.

\bibitem{enns2006netconf}
R.~Enns et~al.
\newblock Netconf configuration protocol.
\newblock Technical report, RFC 4741, December, 2006.

\bibitem{DBLP:journals/pvldb/EskandarianZ19}
S.~Eskandarian et~al.
\newblock Oblidb: Oblivious query processing for secure databases.
\newblock {\em Proc. {VLDB} Endow.}, 13(2):169--183, 2019.

\bibitem{DBLP:journals/jcs/FuhryBBHKS18}
B.~Fuhry et~al.
\newblock Hardidx: Practical and secure index with {SGX} in a malicious
  environment.
\newblock {\em Journal of Computer Security}, 26(5):677--706, 2018.

\bibitem{DBLP:journals/corr/abs-2002-05097}
B.~Fuhry et~al.
\newblock Encdbdb: Searchable encrypted, fast, compressed, in-memory database
  using enclaves.
\newblock {\em CoRR}, abs/2002.05097, 2020.

\bibitem{DBLP:conf/icalp/GellesOW12}
R.~Gelles et~al.
\newblock Multiparty proximity testing with dishonest majority from equality
  testing.
\newblock In {\em {ICALP}}, pages 537--548, 2012.

\bibitem{gerhards2009rfc}
R.~Gerhards et~al.
\newblock Rfc 5424: The syslog protocol.
\newblock {\em Request for Comments, IETF}, 2009.

\bibitem{DBLP:journals/cacm/GoldschlagRS99}
D.~M. Goldschlag et~al.
\newblock Onion routing.
\newblock {\em Commun. {ACM}}, 42(2):39--41, 1999.

\bibitem{DBLP:journals/jcss/GoldwasserM84}
S.~Goldwasser and S.~Micali.
\newblock Probabilistic encryption.
\newblock {\em J. Comput. Syst. Sci.}, 28(2):270--299, 1984.

\bibitem{DBLP:conf/eurosec/GotzfriedESM17}
J.~G{\"{o}}tzfried et~al.
\newblock Cache attacks on {Intel} {SGX}.
\newblock In {\em {EUROSEC}}, pages 2:1--2:6, 2017.

\bibitem{DBLP:conf/icde/HacigumusMI02}
H.~Hacig{\"{u}}m{\"{u}}s et~al.
\newblock Providing database as a service.
\newblock In {\em {ICDE}}, pages 29--38, 2002.

\bibitem{DBLP:journals/corr/abs-2004-05251}
A.~Hekmati et~al.
\newblock {CONTAIN:} privacy-oriented contact tracing protocols for epidemics.
\newblock {\em CoRR}, abs/2004.05251, 2020.

\bibitem{DBLP:conf/ctrsa/IshaiKLO16}
Y.~Ishai et~al.
\newblock Private large-scale databases with distributed searchable symmetric
  encryption.
\newblock In {\em {RSA}}, pages 90--107, 2016.

\bibitem{kjaergaard2012challenges}
M.~B. Kj{\ae}rgaard et~al.
\newblock Challenges for social sensing using wifi signals.
\newblock In {\em Workshop on Mobile systems for computational social science},
  pages 17--21, 2012.

\bibitem{DBLP:conf/huc/KrummH04}
J.~Krumm et~al.
\newblock The nearme wireless proximity server.
\newblock In {\em UbiComp}, pages 283--300, 2004.

\bibitem{krumm2009proximity}
J.~C. Krumm et~al.
\newblock Proximity detection using wireless signal strengths, Mar.~24 2009.
\newblock US Patent 7,509,131.

\bibitem{DBLP:journals/pvldb/LiLWB14}
R.~Li et~al.
\newblock Fast range query processing with strong privacy protection for cloud
  computing.
\newblock {\em {PVLDB}}, 7(14):1953--1964, 2014.

\bibitem{DBLP:conf/icde/LiL17}
R.~Li et~al.
\newblock Adaptively secure conjunctive query processing over encrypted data
  for cloud computing.
\newblock In {\em {ICDE}}, pages 697--708, 2017.

\bibitem{DBLP:conf/wimob/MaierSD15}
M.~Maier et~al.
\newblock Probetags: Privacy-preserving proximity detection using wi-fi
  management frames.
\newblock In {\em {WiMob}}, pages 756--763, 2015.

\bibitem{DBLP:conf/percom/MehrotraKVR16}
S.~Mehrotra et~al.
\newblock {TIPPERS:} {A} privacy cognizant iot environment.
\newblock In {\em {PerCom W}}, pages 1--6, 2016.

\bibitem{meunier2004peer}
J.-L. Meunier.
\newblock Peer-to-peer determination of proximity using wireless network data.
\newblock In {\em {PerComW}}, pages 70--74, 2004.

\bibitem{DBLP:conf/mobisys/PrasadK17}
A.~Prasad and D.~Kotz.
\newblock {ENACT:} encounter-based architecture for contact tracing.
\newblock In {\em WPA@MobiSys}, pages 37--42, 2017.

\bibitem{DBLP:journals/corr/abs-1803-09007}
L.~Radaelli et~al.
\newblock Quantifying surveillance in the networked age: Node-based intrusions
  and group privacy.
\newblock {\em CoRR}, abs/1803.09007, 2018.

\bibitem{DBLP:journals/imwut/SapiezynskiSWLL17}
P.~Sapiezynski et~al.
\newblock Inferring person-to-person proximity using wifi signals.
\newblock {\em {IMWUT}}, 1(2):24:1--24:20, 2017.

\bibitem{schlener2001flexible}
C.~Schlener and S.~Vasudev.
\newblock Flexible snmp trap mechanism, Jan.~30 2001.
\newblock US Patent 6,182,157.

\bibitem{DBLP:journals/cacm/Shamir79}
A.~Shamir.
\newblock How to share a secret.
\newblock {\em Commun. {ACM}}, 22(11):612--613, 1979.

\bibitem{DBLP:conf/sp/SongWP00}
D.~X. Song et~al.
\newblock Practical techniques for searches on encrypted data.
\newblock In {\em {SP}}, pages 44--55, 2000.

\bibitem{DBLP:journals/corr/abs-2004-06818}
Q.~Tang.
\newblock Privacy-preserving contact tracing: current solutions and open
  questions.
\newblock {\em CoRR}, abs/2004.06818, 2020.

\bibitem{epfl}
C.~Troncoso et~al.
\newblock Decentralized privacy-preserving proximity tracing overview of data
  protection and security.
\newblock 2020.
\newblock Available at: \url{https://github.com/DP-3T/documents}.

\bibitem{DBLP:conf/icdcs/WangCLRL10}
C.~Wang et~al.
\newblock Secure ranked keyword search over encrypted cloud data.
\newblock In {\em {ICDCS}}, pages 253--262, 2010.

\bibitem{DBLP:conf/ccs/WangCPZWBTG17}
W.~Wang et~al.
\newblock Leaky cauldron on the dark land: Understanding memory side-channel
  hazards in {SGX}.
\newblock In {\em {CCS}}, pages 2421--2434, 2017.

\bibitem{ylonen2006secure}
T.~Ylonen et~al.
\newblock The secure shell (ssh) protocol architecture, 2006.

\bibitem{DBLP:conf/ccs/YuWRL10}
S.~Yu et~al.
\newblock Attribute based data sharing with attribute revocation.
\newblock In {\em {ASIACCS}}, pages 261--270, 2010.

\bibitem{opaque}
W.~Zheng et~al.
\newblock Opaque: An oblivious and encrypted distributed analytics platform.
\newblock In {\em {NSDI}}, pages 283--298, 2017.

\bibitem{DBLP:conf/huc/ZhouMZSPM16}
M.~Zhou et~al.
\newblock {EDUM:} classroom education measurements via large-scale wifi
  networks.
\newblock In {\em {UbiComp}}, pages 316--327, 2016.

\end{thebibliography}

\appendix

\section{Security Property for Access-Pattern-Revealing Solutions}
\label{app_sec:Security Property for Access-Pattern-Revealing Solutions}
In order to define security property of \textsc{cQuest}, we follow the standard security definitions of symmetric searchable encryption techniques~\cite{DBLP:journals/jcs/CurtmolaGKO11} that define the security in terms of leakages: \emph{setup leakage} $\mathcal{L}_s$ (that includes the leakages from the encrypted database size and leakages from metadata size) and \emph{query leakage} $\mathcal{L}_q$ (that includes search-patterns (\textit{i}.\textit{e}., revealing if and when a query is executed) and access-patterns (\textit{i}.\textit{e}., revealing which tuples are retrieved to answer a query)). Based on these leakages, the security notion provides a guarantees that an encrypted database reveals no other information about the data beyond leakages $\mathcal{L}_s$ and $\mathcal{L}_q$.

Now, before defining security property, we need to formally define \textsc{cQuest}'s query execution method that contains the following three algorithms:

\begin{enumerate}[noitemsep,leftmargin=0.01in]
\item  \textbf{${(K,\mathfrak{R}))\leftarrow \mathit{Setup}(1^k,R)}$:} is a probabilistic algorithm that takes as input a security parameter $1^k$ and a relation $R$. It outputs a secret key $K$ and an encrypted relation $\mathfrak{R}$. This algorithm (as given in Algorithm~\ref{alg:Data Encryption Algorithm}) is executed at \textsc{Quest}'s encrypter, before outsourcing a relation to the cloud.

\item  \textbf{${\mathit{trapdoor}\{1, \ldots, q\} \leftarrow \mathit{Trapdoor\_Gen}(K, \mathit{query})}$:} is a deterministic algorithm that takes as input the secret key $K$ and a query predicate $\mathit{query}$, and outputs a set of query trapdoors, denoted by $\mathit{trapdoor}\{1,\ldots,q\}$. This algorithm (as given in Algorithm~\ref{alg:Query Execution Algorithm}) is executed at \textsc{Quest}'s trapdoor generator and $\mathit{trapdoor}\{1,\ldots,q\}$ are sent to the server to retrieve the desired tuples.

\item  \textbf{${\mathit{results}\leftarrow \mathit{Query\_Exe}(\mathit{trapdoor}\{1,\ldots, q\},\mathfrak{R})}$:} is a deterministic algorithm and executed at the server. It takes the encrypted relation $\mathfrak{R}$ and the encrypted query trapdoors $\mathit{trapdoor}\{1,\ldots, q\}$ as the inputs. Based on the inputs, it produces the results.
\end{enumerate}

\smallskip

In order to define the security notion, we adopt the real and ideal game model security definition~\cite{DBLP:journals/jcs/CurtmolaGKO11}. Based on this game, what the security property is provided is known as indistinguishability under chosen-keyword attack (IND-CKA) model~\cite{DBLP:journals/jcs/CurtmolaGKO11}. IND-CKA prevents an adversary from deducing the cleartext values of data from the encrypted relation or from the query execution, except what is already known.

\medskip
\noindent\textbf{Security Definition.}

Let $\Psi = (\mathit{Setup}, \mathit{Trapdoor\_Gen},\mathit{Query\_Exe})$ be a tuple of algorithms. Let $\mathcal{A}$ be an adversary. Let $\mathcal{L}_s$ be the setup leakage, and let $\mathcal{L}_q$ be the query leakage.

\begin{itemize}[noitemsep,leftmargin=0.01in]
\item
\textbf{${\mathit{Real}_{\Psi,\mathcal{A}}(k)}$:} The adversary produces a relation $R$ and sends it to a simulator. The simulator runs $\mathit{Setup}$ algorithm and produces an encrypted relations $\mathfrak{R}$ that is sent to $\mathcal{A}$. The adversary $\mathcal{A}$ executes a polynomial number of queries on the encrypted relations $\mathfrak{R}$ by asking trapdoors for each of the queries from the simulator. Then, the adversary $\mathcal{A}$  executes queries using $\mathit{Query\_Exe}()$ algorithm and produces a bit $b$.

\item
\textbf{$\mathit{Ideal}_{\Psi,\mathcal{A}}(k)$:} The adversary $\mathcal{A}$ produces a relation $R^{\prime}$. Note that this relation may or may not be identical to the relation $R$, produced in $\mathit{Real}_{\Psi,\mathcal{A}}(k)$. However, $\mathcal{L}_s$ in the ideal world should be identical to the real world. The simulator has neither access to the real dataset $R$, nor access to the real queries. Instead, the simulator has, only, access to $\mathcal{L}_s$ and $\mathcal{L}_q$. The simulator simulates $\mathit{Setup}$ and $\mathit{Trapdoor\_Gen}$ algorithms. Given $\mathcal{L}_s$ and $\mathcal{L}_q$, the simulator produces an encrypted relation $\mathfrak{R}^{\prime}$ and the trapdoors for all queries that were previously executed. The adversary executes the queries and produces a bit $b$.
\end{itemize}

We say $\Psi$ is $(\mathcal{L}_s,\mathcal{L}_q)$-secure against non-adaptive adversary, iff for any probabilistic polynomial time (PPT) adversary $\mathcal{A}$, there exists a PPT simulator such that:
$|\mathit{Pr}[\mathit{Real}_{\Psi,\mathcal{A}(k)} = 1] - [\mathit{Pr}[\mathit{Ideal}_{\Psi,\mathcal{A}(k)} = 1]| \leq \mathit{negl}(k)$, where $\mathit{negl}()$ is a negligible function.

\medskip

The above real-ideal game provides the following intuition: an adversary  selects two different relations, $R_1$ and $R_2$, having an identical  number of attributes and an identical number of tuples. Relations $R_1$ and $R_2$ may or may not overlap. 
The simulator simulates the role of \textsc{Quest} encrypter to produce an encrypted relation and provides it to the adversary. On the encrypted data, the adversary executes a polynomial number of queries. The adversarial task is to find the relation encrypted by the simulator, based on the query execution. The adversary cannot differentiate between the two encrypted relations, since if the adversary cannot find which encrypted relation is produced by the simulator with probability non-negligibly different from 1/2, then the query execution reveals nothing about the relation.

\end{document}